\documentclass[11pt,oneside]{book}

\usepackage{hyperref}

\usepackage{amsmath,amsfonts,amssymb,amsthm}
\usepackage{latexsym}
\usepackage{pifont}
\usepackage{slashed}
\usepackage{graphicx,chngcntr,vmargin,booktabs,caption}
\usepackage{fancyhdr,bigints,cancel}
\usepackage{enumitem}
\usepackage{subfigure}
\usepackage{braket}
\usepackage{epsf}
\usepackage{mathrsfs}
\usepackage{cite}
\usepackage[usenames,dvipsnames]{color}
\usepackage{pdfpages}

\setmarginsrb{25mm}{20mm}{25mm}{20mm}{10mm}{10mm}{10mm}{10mm}
\newcommand{\arxivlink}[1]{\href{http://arxiv.org/abs/arXiv:#1}{\textcolor{blue}{arXiv:#1}}}

\newcommand{\oo}{\infty}
\newcommand{\md}{\mathrm{d}}
\newcommand{\be}{\begin{equation}}
\newcommand{\ee}{\end{equation}}
\newcommand{\bea}{\begin{eqnarray}}
\newcommand{\eea}{\end{eqnarray}}

\renewcommand{\d}{\mathrm{d}}

\newcommand{\as}{\alpha_s}
\newcommand{\Bor}{\mathrm{B}}
\newcommand{\abs}[1]{\left| #1 \right|}
\newcommand{\Lum}{\mathscr{L}}
\newcommand{\plus}[1]{\left(#1\right)_+}

\numberwithin{equation}{section}

\begin{document}

\includepdf[pages={1}]{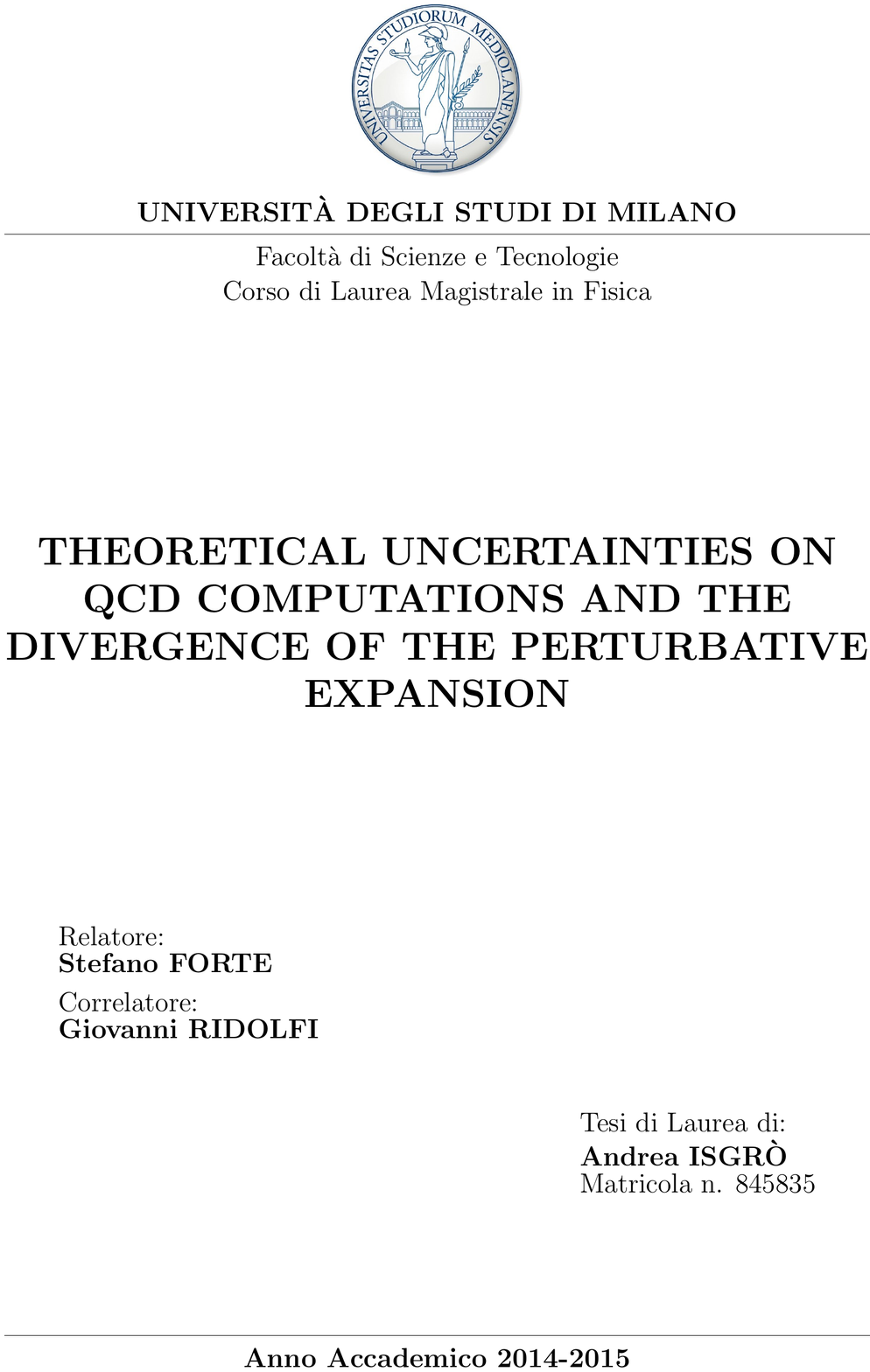}
\phantomsection \chapter*{Summary} 
\thispagestyle{empty} 

The purpose of this thesis is to analyze the divergent behaviour of perturbative series in QCD, in order to estimate the uncertainty on finite-order perturbative results. With a new LHC era starting this year (2015), precision in theoretical predictions is important more than ever, especially for the newly discovered Higgs boson. Unluckily, the perturbative series of the cross section of Higgs production in gluon fusion (the most relevant channel at the LHC) converges rather slowly. Furthermore, the methods we have to determine the size of higher order corrections have a poor accuracy, and most of them lack of a physical interpretation.

Our starting point is that any observable in QCD or QED which is a perturbative expansion in the coupling constant $\as$ or $\alpha$ is divergent, and this has been known since a long time (Dyson, 1952). In Chapter~\ref{ch:divergentseries} we will first illustrate Dyson's argument. Then, we will introduce some tools that can help us deal with divergent series, in particular Borel summation, and see some instructive examples. Finally, we will talk about the three most important known sources of divergence in QCD: the Landau pole, renormalons and instantons.

The Landau pole divergence is closely related to the concept of resummation, which we shall introduce in Chapter~\ref{ch:landaupole}. A certain class of corrections to the leading order cross section, the emission of soft gluons, is known to produce logarithmically enhanced contributions, which can be resummed to all orders. However, this sum is convergent in Mellin space, but cannot be transformed back in physical space, due to a branch cut for large values of the Mellin moment $N$. The presence of the Landau pole is ultimately responsible for this branch cut. As a result, the expansion of the resummed cross section in physical space, which happens to be (at least for the Higgs case) a very good approximation to the exact fixed-order calculation, is a divergent series.

In Chapter~\ref{ch:renormalons}, we will see that another class of perturbative corrections, the fermion bubble diagrams, produce a factorial divergence when integrating over very high (UV) or very low (IR) internal momentum. This divergence is called renormalon divergence and we will see how one can generalize the concept of renormalon, defining it as a pole in the real axis of the Borel plane, located at an integer multiple of $\beta_0$.

There are other poles in the Borel integration path: the instantons. They produce another factorial divergence which is this time related to the growing number of Feynman diagrams with increasing perturbative order. In some quantum mechanical systems and even in some quantum field theories, one can sum the instanton-anti instanton contribution to the divergent perturbative expansion and obtain a so-called resurgent trans-series, eliminating completely the divergence. Unluckily, this is not the case of QCD, where IR renormalons are much closer to zero (and therefore lead to a faster divergence) than instantons.

After introducing the sources of divergence, in Chapter~\ref{ch:theorunc} we will illustrate the current most used models for theoretical uncertainties. Since most of the theoretically calculable cross sections are known up to the third or fourth perturbative order at best, those methods make an attempt to describe the effect of the unknown higher orders.
An issue which we will not discuss in detail but deserves a special mention is the impact of parton distribution functions (PDFs) on theoretical uncertainties.

Finally, in Chapter~\ref{ch:results} we will derive our own model for estimating theoretical uncertainties, based on what we know about the sources of divergence mentioned above. We will then apply that model to 
two of the most relevant processes at the LHC: 
 Higgs production 
and $t \bar t$ production. 
Our first concern will be to compute the perturbative order at which each of the sources of divergence starts to kick in. After that, we will compute our estimate for the theoretical uncertainties on the last known order and we will compare our result with that of the already known models.

\newpage \thispagestyle{empty}
\tableofcontents \thispagestyle{empty}
\newpage \thispagestyle{empty}
\chapter{Divergent series in QCD}
\thispagestyle{empty}
\label{ch:divergentseries}
The Large Hadron Collider has just restarted (early 2015) after two years of maintenance and upgrading. As we write, the center of mass energy of the collisions is set to 13 TeV, which means that the cross sections of most of the events that we can observe are significantly increasing with respect to a couple of years ago. With no evidence of Beyond the Standard Model physics at 8 TeV, precision measurements in the SM are needed from an experimental and a theoretical point of view.

Let us focus on theoretical accuracy, starting with an example. The cross section for the production of a Higgs boson in gluon fusion at LHC @ 13 TeV is, as any observable computed in quantum field theory, a perturbative series. This means that if we want to give a theoretical prediction for the cross section, its form will be
\begin{equation}
\label{eq:higgsxs}
\sigma_{gg \to H} = c_0 + c_1 \as + c_2 \as^2 + \dots,
\end{equation}
where $\as$ is the strong coupling constant, which computed at the Higgs mass scale has the value~$\as (m_H) = 0.1126$. Being $\as$ a perturbative parameter, one could expect that the difference between, for example, the Next to Leading Order (NLO) and the NNLO should be small. If this were case, the calculation of high perturbative orders, which usually requires a lot of time and effort, would be only needed if one wanted to have an extraordinary precision. However, as we can see in Table~\ref{tab:higgsxs}, this perturbative series is slowly convergent at the first known orders. The perturbative corrections can even be bigger than the whole cross section at the previous order. The situation does not improve changing the energy of the collision.
\begin{table}[b] 
\centering
\begin{tabular}[c]{rccccc}
c.m.e     &LO [pb] &NLO [pb] &NNLO [pb]& $\frac{\text{NLO-LO}}{\text{LO}}$ \% & $\frac{\text{NNLO-NLO}}{\text{NLO}}$ \% \\
 \midrule
8 TeV    & 5.37      & 12.78     & 17.15       & 138.18 \%                                          & 34.21 \% \\
13 TeV  & 12.25    & 29.40     & 39.24       & 139.98 \%                                          & 33.46 \%   \\
\end{tabular}
\caption{Higgs cross section in gluon fusion at 8 TeV and at 13 TeV}
\label{tab:higgsxs}
\end{table}
This is one issue we want to address: if we want to give a good theoretical prediction to the Higgs cross section, we need to take into account in some ways even the unknown higher orders.

Another important issue is the fact that series like~\eqref{eq:higgsxs} are known to be slowly convergent. Actually, Dyson in 1952~\cite{dyson} showed that any perturbative expansion in QCD is divergent. As a matter of fact, we know well some physical sources of divergence: instantons, renormalons and the Landau pole in soft gluon resummation. What we do not know is the perturbative order at which these divergences start to kick in. If, for example, the series~\eqref{eq:higgsxs} started to diverge at the 4$^\text{th}$ perturbative order, then any exact N$^4$LO calculation would not make sense, because that result would be distant from the exact value of the series.

This chapter is so organised: first we will present Dyson's argument and show why all perturbative series in QED and QCD are divergent. Then we will describe the basic tools we can use to treat divergent series.

\section{Dyson's argument}
In 1952, Freeman Dyson presented a very straightforward argument~\cite{dyson} which led to the conclusion that all the power series expansions in use in quantum electrodynamics are divergent after the renormalization of mass and charge. This same argument can be applied with just a little effort to QCD.

The starting point is that all existing methods of handling problems in quantum electrodynamics give results in the form of power series in $\alpha = e^2/(4 \pi)$. Let us consider a generic observable $\sigma$:
\begin{equation}
\sigma(e^2) = a_0 + a_2 e^2 + a_4 e^4 + \dots.
\end{equation}
Of course, the coefficients $a_i$ are finite after mass and charge renormalization. The series as a whole, though, cannot be treated with the same techniques. If the series converges, its sum is a calculable physical quantity. But if the series diverges,
it becomes difficult to calculate or even of define the quantity which is supposed to be represented by the series.

We know that QED is equivalent to a theory of the motion of charges acting on each other by a direct action at a distance. The interaction between two like charges is proportional to $e^2$. Suppose now that the series $\sigma(e^2)$ converges for some positive value of $e^2$, this implies that $\sigma(e^2)$ is an analytic function of $e$ at $e=0$. Then, for small values of $e$, $\sigma(-e^2)$ will also be an analytic function with a convergent power series expansion.

But we can also find a physical interpretation for $\sigma(-e^2)$: it is the value that would be
obtained for $\sigma$ if the interaction
between charges of the same type had a minus sign. In a fictitious world like that, charges of the same type attract each other and the classical macroscopic potential is just the Coulomb potential with the sign reversed. But in these conditions the vacuum state is not the state of lowest energy. In fact, one could construct a pathological state by creating a large number of electron-positron pairs and bringing the electrons and the positrons in two separate regions.  In a state thus made, the negative potential energy of the Coulomb forces is greater than the sum of the total rest energy and kinetic energy of the particles. 

This can be done without
using particularly small regions or high charge densities,
so that the validity of the classical Coulomb potential
is not in doubt. Let us suppose now that a system is given at a certain time with only a few particles present. There exists a high potential barrier separating this physical state from the pathological state described above. However, due to the quantum mechanical tunnel effect, there is a finite probability that the system will evolve towards the pathological state. Therefore, any physical state is unstable against the spontaneous creation of infinite particles.  Furthermore, once a system finds itself in a pathological state, there will be an inevitable creation of
more and more particles. In these 
conditions it is impossible that the integration of the
equations of motion, starting from a given state of the
fictitious world, should lead to well defined analytic functions. Therefore $\sigma(-e^2)$ cannot be analytic and the 
the series cannot be convergent.

\section{Divergent series}
We have just seen that any perturbative series in QED or QCD is intrinsically divergent. Even so, there is much that we can say about divergent series and the interpretation of their sum. An exhaustive review about the treatment of divergent series can be found in Ref.~\cite{hardy}, we will mainly follow Appendix D of Ref.~\cite{bonvinithesis} for our purposes. Let's start with some definitions.
A generic series
\be
\label{eq:genericseries}
S = \sum_{k=0}^\infty c_k
\ee
is \emph{convergent} if, being its partial sums
\be
s_n = \sum_k^n c_k,
\ee
the following limit
\be
s = \lim_{n\to\infty} s_n
\ee
is finite. In this case such limit $s$ is called the sum of the series and we say that $S=s$.
Otherwise, we say that the series is \emph{divergent}. Another very useful definition is that of absolute convergence: a series is  \emph{absolutely convergent} if the series of the absolute values $\sum_k |c_k|$ is convergent.

If the series is a power series,
\be
S(z) = \sum_k c_k z^k,
\ee
then its convergence depends on the value of $z$. In particular, what happens is that the series is convergent for values of $z$ that lie in a circle of radius $r$, where
\be
r = \lim_{k\to\infty} \abs{\frac{c_k}{c_{k+1}}}.
\ee
The convergence of the series implies the analyticity of the sum $s(z)$ inside such circle. Therefore, if we expand a function $f(z)$ around some point $z_0$,
the radius of convergence of the expansion can be at most the distance between
$z_0$ and the singularity that is closest to it. 
\subsection{Asymptotic expansions}
Sometimes, a function $f(z)$ might admit a series expansion around a certain value of $z$ that is a divergent series.
In other words, if we expand $f(z)$ around $z=0$, we have that a series expansion
\be
S(z) = \sum_k c_k z^k
\ee
is  \emph{asymptotic} to $f(z)$ if
there exists a constant $K$ such that
\be
\abs{f(z) - s_n(z)} \leq K\, c_{k+1} \abs{z}^{k+1}
\ee
for all $n$.
\subsection{Sum of divergent series}
\label{sec:divergent_series}
Whenever we have a divergent series, by definition the limit of the partial sum is infinite (or does not exists). However, this has nothing to do with the finiteness of the sum of the series. In fact, we can define the sum of a series in another way, it doesn't have to be the limit of the sequence of the partial sums.

As an example, let's consider the following (divergent) series
\be \label{eq:111}
P = \sum_{k=0}^\infty (-1)^k = 1-1+1-1+\ldots.
\ee
This series is divergent because the partial sums
\be
p_n = \sum_{k=0}^n (-1)^k =
\begin{cases}
0 & \text{for even $n$}\\
1 & \text{for odd $n$}
\end{cases}
\ee
oscillate between $0$ and $1$, and therefore the limit for $n\to\infty$ of $p_n$ is not defined.
However, there are multiple arguments that assign to the sum the value $1/2$.
For example, we can manipulate the definition of the series to obtain an equation for $P$:
\bea
  P = \sum_{k=0}^\infty (-1)^k =1 + \sum_{k=1}^\infty (-1)^k 
  = 1 + \sum_{k=0}^\infty (-1)^{k+1} 
  = 1 - \sum_{k=0}^\infty (-1)^k 
 = 1-P
\eea
from which we get $P=1/2$.
Another way to obtain $P=1/2$ is to consider the power series
\be\label{eq:111_z}
P(z) = \sum_{k=0}^\infty (-z)^k
\ee
which is  convergent in the circle $|z|<1$ in the complex plane. In that region, the sum is
\be
P(z) = \frac{1}{1+z}.
\ee
We can analytically extend the function $P(z)$ to the entire complex plane, apart from $z=-1$.
The starting series \eqref{eq:111} is obtained when $z=1$, outside the convergence domain, but with analytical continuation we can assign to the sum the value  $P=P(1)=1/2$.
\subsection{Borel summation}
\label{sec:Borel}
Once that we have a divergent series, there are several ways to assign a value to its sum.
In the following, we will concentrate on one method: the so-called Borel method.

Let us define the Borel transform of a generic series, like that in equation~\eqref{eq:genericseries}, as
\begin{equation}
\label{eq:boreltransform}
\Bor[S](t) = \sum_{k=0}^\infty \frac{c_k}{k!} \, t^k.
\end{equation}
In the Borel method, we define the sum of the series as
\be
\tilde S  = \int_0^\infty \d t \, e^{-t} \, \Bor[S](t)
\ee
If the series $S$ were convergent, we could exchange the sum with the integral and integrate term by term. By doing this, we would obtain that the sum of the series would be exactly $\tilde S$.

We say that a series is Borel-summable $\Bor$-summable  if
\begin{itemize}
\item  its Borel transform converges $\forall t$
\item  $\Bor[S] (t)$ is defined on $0\leq t \leq \infty$
\item  the integral converges.
\end{itemize}

We can generalize the Borel method and define higher-order Borel transform:
\be
\tilde S_n = \int_0^\infty \d t_1  \int_0^\infty \d t_2 \cdots \int_0^\infty \d t_n \, e^{-(t_1+ t_2 + \dots +t_n)}  \Bor_n[S] (t_1, t_2, \dots, t_n),
\ee
where
\be
\Bor_n[S] (t_1, t_2, \dots, t_n) = \sum_{k=0}^{\infty} \frac{c_k}{(k!)^n}(t_1 t_2\cdots t_n)^k
\ee
is the $n$-Borel transform.
It can be shown that if a series is $\Bor_n$-summable it is also $\Bor_k$-summable, $\forall k>n$.
\subsection{Examples of Borel summation}
\label{sec:borelsummationexamples}
Once that we have defined the procedure of Borel summation, it is instructive to try and apply this method to some known divergent series. As we will see, most of the series that we are about to study have a physical meaning and correspond to a specific source of divergence.
\subsubsection{Alternating series}
The first example is just a simple exercise that confirms the results obtained for the previously cited series Eq.~\eqref{eq:111}.
The Borel transform of the series is
\be
\Bor[P] (t) = \sum_{k=0}^\infty \frac{(-t)^k}{k!} = e^{-t}.
\ee
The series converge in the whole complex plane, and the Borel sum is
\be
\tilde P = \int_0^\infty \d t\, e^{-t} e^{-t} = \frac12,
\ee
like we previously found.

\subsubsection{Factorial divergence, alternating sign}
Consider the following series which, as we shall see, represents the divergent behaviour induced by the infrared renormalons:
\be
R_\text{IR} = \sum_{k=0}^\infty (-1)^k k!.
\ee
Its Borel transform reads
\be
\Bor[R_\text{IR}](t) = \sum_{k=0}^\infty (-t)^k = \frac{1}{1+t},
\ee
where the convergence radius of $\Bor_n$ for $n=1$ is $|t|<1$,
while it can be shown that the $n>1$ Borel transforms have infinite convergence radius.
The Borel sum is given by
\be
\tilde R_\text{IR} = \int_0^\infty \d t\,e^{-t} \frac{1}{1+t}.
\ee
The result is the same at all orders
\be
\tilde R_\text{IR} = -e\,\text{Ei}(-1) = 0.596347.
\ee
The function Ei$(x)$ is called \emph{Exponential integral} and is a special function defined on the complex plane as
\begin{equation}
\text{Ei}(x) = - \int_{-x}^{+\infty} \d t \, \frac{e^{-t}}{t}
\end{equation}
The definition above can be used for positive values of $x$, but the integral has to be understood in terms of the Cauchy principal value due to the singularity of the integrand at zero.

\subsubsection{Factorial divergence, same sign}

Consider now the divergent series, similar to that of the previous example and physically related to the behaviour of ultraviolet renormalons:
\be
R_\text{UV}=\sum_{k=0}^\infty k!.
\ee
The corresponding Borel transform is
\be
\Bor[R_\text{UV}] (t) = \frac{1}{1-t}.
\ee
Once again, the first order Borel transform has convergence radius $|t|<1$,
and higher order transforms converge everywhere.
However, for this series the Borel inversion intergal does not converge because of a pole ($t=1$) in the integration path for $n=1$ and because of the bad behaviour at $t\to\infty$ for the higher-order Borel transforms.
However, the first order Borel integral
\be
\tilde R_\text{UV} = \int_0^\infty \d t\,e^{-t} \frac{1}{1-t}
\ee
can still have a meaning if we deform
the integration contour in the complex $t$-plane and avoid the pole.
But, for this reason, the result has an ambiguity, given by the two possible
way of avoiding the pole (above or below the positive real axis).
The result is 
\be\label{eq:borel_sum_series_k!_1}
\tilde R_\text{UV} =  \frac 1e \left[ \text{Ei}(1) \pm i\pi \right].
\ee
However, if we tried to use any of the $n>1$ methods, we would end up with a different result. In this case, the only way to determine which result is correct is to consider the power series
\begin{equation}
R(z) = \sum_{k=0}^\infty (-z)^k \, k!.
\end{equation}
\section{Known sources of divergence}
\label{sec:knownsources}
We have seen that perturbative series in QCD are always divergent. We have also introduced some tools that can help us deal with divergent series. What is left to see is how the divergence arises in computations. There are three known sources of divergence in QCD that we are going to explore in this thesis:
\begin{itemize}
\item Landau pole divergence
\item Renormalons
\item Instantons
\end{itemize}

The origin of the \emph{Landau pole divergence} resides in soft gluon resummation. The basic idea is that
 soft gluon radiation has the effect of replacing the hard scale by a softer scale that is related to the process of radiation.
For example, let us assume that a physical process is
characterized by the hard scale $Q^2$ and a generic scaling variable $0\le
z\le 1$. What happens is that near the $z=1$ region, close to threshold,  the
resummation of large logs of $1-z$ replaces $\as(Q^2)$ with $\as(Q^2(1-z))$.
Since $\as$ has a rescaled argument, it becomes too large to be treated with the perturbative approach. The physical interpretation of this fact is that  when $z\to1$, the
center-of-mass energy is just sufficient to produce the given final
state, so in
this limit the process becomes elastic.

In practice,  at some low scale
$\Lambda$ (the position of the Landau pole) the strong coupling explodes, so when
\be
z=z_L\equiv 1- \frac{\Lambda^2}{Q^2}\label{eq:xldef}
\ee
resummed results become meaningless. The scale $\Lambda$ is usually
taken to be $\Lambda_{\text QCD}$.

It is usually easier to consider resummed results in 
terms of the variable $N$ which is
Mellin conjugate to $z$. The $z \to 1$ region corresponds to  
$N\to\infty$. If the $N$--space resummed result is expanded
perturbatively in powers of $\alpha_s(Q^2)$ and then Mellin transformed
back to $z$ order by order, one ends up with a divergent series, and the source of this divergence is the presence of the Landau pole.
\\

\emph{Renormalons} will be discussed in Chapter~\ref{ch:renormalons}. Firstly identified as infinite chains of bubble diagrams, the renormalons are singularities in the Borel complex plane. They are related to small and large momentum behaviour of perturbative corrections, and  divided respectively into \emph{Infrared (IR) renormalons} and \emph{Ultraviolet (UV) renormalons}. The divergent behaviour induced by renormalons strongly depends on the position of the poles. We will see that the Borel transform Eq.~\eqref{eq:boreltransform} of a perturbative series that has the renormalon problem presents poles at $t = m \beta_0$, with $m$ an integer and $\beta_0$ the first term of the $\beta$ function. In particular, the closer the pole is to 0, the sooner (in terms of perturbative orders) the divergence occurs. There will then be a leading renormalon ($m=1$) that will dictate the divergent behaviour of the series.

Furthermore, we will see that renormalons are connected with the concept of \emph{power corrections}. The fact that there are poles in the integration path of the Borel integral Eq~\eqref{eq:borelintegral}, leads to an ambiguity in the results. This ambiguity can be interpreted in terms of the difference between the asymptotic value of the series and the real sum. We will see how these power correction vary with the scale and how this scaling behaviour is related to the position of the leading renormalon.
\\

\emph{Instantons} are field configurations fulfilling the classical equations of motion in Euclidean spacetime, which can be interpreted as a tunneling effect between different topological vacua. In quantum mechanics there are cases, e.g., the double-well potential, in which the ambiguity introduced by the Borel summation can be cured by a procedure called the Bogomolny-Zinn-Justin (BZJ) prescription. Here the perturbative ambiguity cancels against a non-perturbative contribution from instanton--anti-instanton events. The sum of the perturbative and non-perturbative semiclassical expansions in quantum mechanics apparently produces ambiguity free (and accurate) results. 

One may ask if this idea can work in field theory, and in particular in QCD. The answer~\cite{thooft} is that it does not work for gauge theories on $\mathbb{R}^4$ due to the above-mentioned IR renormalon problem. Reference~\cite{unsal} argues that this method does work on $\mathbb{R}^3 \times S^1$ in a gauge theory continuously connected to one on $\mathbb{R}^4$, but the generalization to QCD is an open problem.

\chapter{Divergence in resummed series}
\thispagestyle{empty}
\label{ch:landaupole}

\noindent
In this Chapter we study one of the known sources of divergence in perturbative QCD, the \emph{Landau pole divergence}. First, we will review the basics of soft-gluon
resummation. Our discussion will mainly follow Refs.~\cite{bonvinithesis,rottolithesis}.

We have already anticipated in Section~\ref{sec:knownsources} that, because of the presence of the Landau pole, the inverse Mellin transform of the resummed result in the soft-gluon limit does not exist. If we expand the resummed cross section in powers of $\as$, each of the terms has a finite inverse Mellin transform, but the correspondent perturbative series in physical space is divergent.  We will describe a prescription to obtain the resummed result in $z$ space, the Borel prescription~\cite{borelprescription}. This is not the only way to deal with the Landau pole divergence: another widely used prescription, which we will not cover here, is the minimal prescription~\cite{minimalprescription}.

Finally, we will show that an approximation can be constructed using the expansion of the resummed cross section in the soft limit. For processes like Higgs production in gluon fusion, this approximation succeeds in predicting fixed order calculations. The divergent behaviour of the expansion of the resummed cross section is related to that of the exact fixed order calculation.

\section{Soft-gluon resummation}
The generic coefficient function $C(z,\as)$ suffers from kinematical enhancements due to gluon emissions when $z \to 1$.  The divergences that appear when calculating virtual contributions are exactly cancelled by those soft gluon emissions. In fact, a generic physical cross section needs to be inclusive over arbitrarily soft particles in the final state, since any detector has a finite energy resolution. Nevertheless, soft-gluon effects can still be large in some kinematic regions. Because of that, calculations to all orders of perturbation theory are necessary to achieve reliable predictions. In this framework, resummation is basically an all-order summation of certain classes of logarithms.

\begin{figure}[t]
  \centering
  \includegraphics[width=0.8\textwidth]{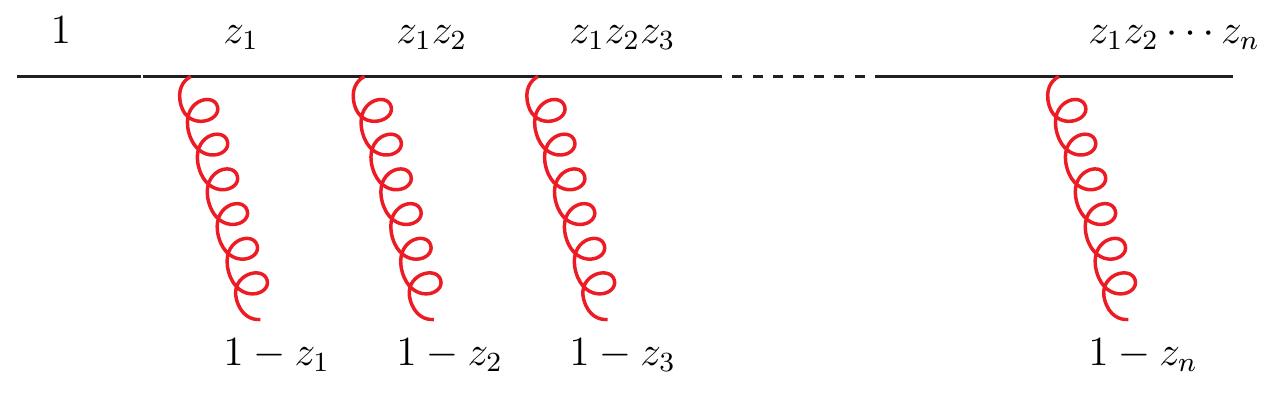}
  \caption{Emission of $n$ gluons from a quark parton line. The quantity $z_i$ represents the energy fraction for each line. Figure taken from Ref.~\cite{bonvinithesis}.}
  \label{fig:soft-gluons}
\end{figure}

To see this in practice, consider as an example the case of a quark parton line emitting $n$ gluons,
as in Fig.~\ref{fig:soft-gluons}.
The energy fraction of each emitted gluon to the quark energy is $1-z_i$. Therefore, after $n$ emissions the energy of the quark will be $z=z_1 z_2\cdots z_n$ times its initial energy. When we integrate over the phase space of the emitted gluons, we have a kinematic enhancement for each gluon. It can be shown that all these terms convert into the sequence
\be
\as^n \left[\frac{\log^k(1-z)}{1-z}\right]_+,\qquad
0\leq k\leq an-1
\label{eq:log(1-z)_tower}
\ee
where $a=1$ for deep inelastic scattering and
$a=2$ for Drell-Yan and Higgs production (we will consider this last case).

It is clear now why we need resummation: there will always be a certain kinematical region where $z$ is such that
\be
\as \log^2(1-\bar z) \sim 1.
\ee
In this case, all terms in the perturbative series are of the same order, and any truncation would neglect a huge part of the result.

\subsection{Resummation in Mellin space}

The concept of resummation has been known for a long time~\cite{sterman,catani,forteresumm}.
This section only serves as a reminder of the most important concepts.

Whenever we try to compute a cross section, we need of course a matrix element and an integral over the phase space. In the case of $n$ gluon emission from a parton line, the calculation of the matrix element can be performed in the eikonal approximation, according to which the matrix element $\mathfrak{M}_n$ factorizes as
\be
\mathfrak{M}_n(z_1,\ldots,z_n) \overset{\rm soft}{\simeq} \frac{1}{n!} \prod_{i=1}^n \mathfrak{M}_1(z_i),
\ee
where $\mathfrak{M}_1$ is the matrix element for the single emission. 
Unluckily, the phase space in physical space is not factorized, because of the Dirac $\delta$ that expresses the conservation of momentum
\be\label{eq:gluon_energy_ps}
\d z_1\, \d z_2\cdots \d z_n \, \delta(z-z_1 z_2\cdots z_n).
\ee
However,  in Mellin space even the phase space factorizes
\be
\int_0^1 \frac{\d z}{z}\, z^N\, \delta(z-z_1 \cdots z_n) 
= z_1^{N-1} \cdots z_n^{N-1}.
\ee

We remind that the Mellin transform of a 
function $f(z)$ is defined as
\bea
f(N) \equiv \mathcal M[f] (N) \equiv \int_0^1 \d z\ z^{N-1} f(z).
\eea
The Mellin transform can be seen as a Laplace transform where a change of variables has 
been performed.
The inverse Mellin transform is
\bea
f(z) = \frac{1}{2 \pi i} \int_{c-i\infty}^{c+i\infty} dN \ z^{-N} f(N),
\eea
where $c$ is greater than the real part of the rightmost singularity. It is easy to show that the Mellin transform of a convolution is the product of the Mellin tranforms
\bea
(f \otimes g)(N) &= \displaystyle \int_0^1 dx\ x^{N-1} \int_x^1
\frac{dy}{y} f(y) g \left(\frac{x}{y} \right) =
\int_0^1 dx\ x^{N-1} \int_0^1 dy \int_0^1 dz f(y) g(z) \delta(x-yz)\nonumber \\
&= \displaystyle \int_0^1 dy \ y^{N-1} f(y) \int_0^1 dz z^{N-1} g(z) = f(N) g(N).
\eea

Back to the emission of multiple gluons, it can be shown that the threshold region $z\sim 1$ corresponds to the region
of large Mellin moment $N$. In particular, the sequence of contributions Eq.~\eqref{eq:log(1-z)_tower}
converts into the tower
\be\label{eq:logN_tower}
\as^n \log^k\frac1N,\qquad
0\leq k\leq 2n.
\ee
Therefore,  the coefficient function in $N$-space at order $\as^n$  is given  by
\be
C^{(n)}(N) \overset{\rm soft}{\simeq} \frac{1}{n!} \left[ C^{(1)}_{\rm soft}(N)\right]^n,
\ee
where $C^{(1)}_{\text soft}(N)$ is the Mellin transform of soft terms up to order $\as$.
If, for example,  we only consider the leading term, the coefficient function becomes
\be\label{eq:C1soft}
C^{(1)}_{\rm soft}(N) = \int_0^1 \d z\, z^{N-1} \,4A_1 \plus{\frac{\log(1-z)}{1-z}}
\overset{N\gg 1}{\simeq} 2A_1\, \log^2 \frac1N
\ee
where $A_1=C_F/\pi$ for the Drell-Yan case and $A_1=C_A/\pi$ for the Higgs case.

Now it comes naturally that the soft terms can be resummed explicitly, and the result is the exponential:
\be\label{eq:exponentiation}
C^{\rm res}(N,\as) = \sum_{n=0}^\infty \as^n \left[C^{(n)}(N)\right]_{\rm soft} = \exp \left[\as\, C^{(1)}_{\rm soft}(N) \right].
\ee
Since we have only considered the first order in $\as$, we say that this result is valid only at leading-logarithmic (LL) accuracy.
In fact, only the highest power $k=2n$ in the tower of logs of Eq.~\eqref{eq:logN_tower} is resummed.
Furthermore, this expression does not take into account the running of $\as$. 
It can be proven that the most general expression for the
$N$-space resummed coefficient function is
\bea\label{eq:Cres}
C^{\textrm{res}}(N, M^2) = {\bar g}_0(\alpha_s) \exp \bar {\mathcal S}\left( M^2, \frac{M^2}{N^2} \right),
\eea
where $ \bar {\mathcal S}$ is called \emph{Sudakov form factor} and is defined as
\bea\label{eq:barS}
\bar {\mathcal S} \left(M^2, \frac{M^2}{N^2} \right) =
\int_0^1 \d z \ z^{N-1} \left[\frac{1}{1-z} \int_{M^2}^{M^2(1-z)^2}
\frac{d \mu^2}{\mu^2} 2 A \left(\alpha_s(\mu^2)\right) + 
D\left( \alpha_s([1-z]^2 M^2)\right) \right]_+.
\eea
We have introduced the functions $\bar g_0(\alpha_s)$, $A(\alpha_s)$ and $D(\alpha_s)$. They are always represented as
power series in $\alpha_s$, with $\bar g_0(\alpha_s) = 1 + \mathcal O(\alpha_s)$.
$D(\alpha_s)$ and $\bar g_0(\alpha_s)$ are process-dependent,
while $A(\alpha_s)$ is process independent: it is the coefficient of the soft singularities
in the Altarelli-Parisi splitting function. 
The resummed coefficient function can be written in a form where the Sudakov exponent is organized in powers of logs and $\as$:
\be\label{eq:cresg0expS}
C^{\textrm{res}}(N, M^2) = g_0(\alpha_s) \exp \mathcal S \left(\bar \alpha
L, \bar \alpha \right)
\ee
\be
 \bar \alpha \equiv 2 \beta_0 \alpha, \qquad L \equiv \log \displaystyle \frac{1}{N}, 
\ee
where $g_0$  can be written as
\bea
g_0(\alpha_s) = 1 + \sum_{j=1}^\infty g_{0j} \alpha_s^j, 
\eea
and $\mathcal S$ has the following logarithmic expansion
\be
\mathcal S(\bar \alpha L, \bar \alpha) = \frac{1}{\bar \alpha} g_1(\bar \alpha L) +
g_2(\bar \alpha L) 
+ \bar \alpha g_3 (\bar \alpha L) + \bar \alpha^2 g_4(\bar \alpha L) + \ldots.
\label{eq:sudakovexp}
\ee
The functions $g_i$ can be obtained performing the integrals Eq.~\eqref{eq:cresg0expS}, and
they are determined (as shown, for example, in Refs.~\cite{bonvinithesis, rottolithesis}) by a small number of coefficients of the expansion of $A(\as)$ and $D(\as)$, and 
are of order $g_1(\bar \alpha L) = \mathcal O(\as^2)$, 
$g_i(\bar \alpha L)= \mathcal O(\as)$ for $i>1$.

One finds that the N$^p$LO anomalous dimension is necessary in order to obtain
$g_{p+1}$, which enters the result at N$^p$LL accuracy. Conversely, most of the times the process-dependent functions are determined by matching the expansion of the resummed result with a fixed order computation.
Finally, predictions for phenomenology at N$^p$LO+N$^k$LL accuracy are obtained by combining the the fixed-order computation with the resummed 
coefficient function expanded in powers of $\as$. In this step it is important to remember to subtract the double-counting terms:
\bea
C^{\textrm{N$^p$LO}}_{\textrm{N$^k$LL}}(N, \as) =
\sum_{j=0}^p \as^j C^{(j)}(N) + C^\textrm{res}_{\textrm{N$^k$LL}} (N, \as) 
-\sum_{j=0}^p \frac{\as^j}{j!} \left[ \frac{d^j C^\textrm{res}_{\textrm{N$^k$LL}}(N, \as)}{d \as^j}\right]_{\as=0}.
\label{eq:matching}
\eea

\section{The Landau pole divergence}
\label{sec:landaupole}
So far we have seen how soft gluon resummation is performed in Mellin space. As we anticipated in Section~\ref{sec:knownsources}, the resummed cross section in Mellin space has a branch cut for large values of $N$ (soft region). This means that if we expand in $\as$ the resummed cross section and then compute the inverse Mellin transform term by term, then each inverse transform exists, but the resulting series is divergent in $z$ space. Let us see how this divergence arises and why it is related to the Landau pole.
Consider a generic observable in physical space
$\sigma(Q^2,z)$ and
its $N$-space transform
\be
\sigma(Q^2,N)=\int_0^1 \d z\,z^{N-1}\,\sigma(Q^2,z).
\ee
Since the resummation has the form of an exponentiation, it easier to work in terms of the physical anomalous dimension, which is defined to be
\be
\label{eq:anomalousdim}
\gamma(\as(Q^2),N)=\frac{\partial\ln\sigma(Q^2,N)}{\partial\ln Q^2}.
\ee
Let us consider for now the structure functions for DIS. Similar observations can be made for most of the other processes that are relevant at the LHC (e.g. DY, Higgs production).
The resummed expression of the physical anomalous dimension $\gamma(\as(Q^2),N)$ has the
form~\cite{forteresumm,catani,sterman}
\be
\gamma(\as(Q^2),N)=
\int_1^N\frac{dn}{n}\,\sum_{k=1}^\infty g_k\,\as^k(Q^2/n)+O(N^0),
\label{eq:gamma}
\ee
where $g_k$ are constants which be determined either from first principles~\cite{catani,sterman} or by comparison with the fixed-order calculations. Here, the sum over $k$ represents the sum over successive orders in logarithmic accuracy. The resummed expression of $\gamma(\as(Q^2),N)$ at N$^k$LL can be used to compute the resummed cross section in Mellin space at N$^k$LL, but then one needs to compute the inverse Mellin transform of this quantity to obtain the resummed cross section in physical space.

The main effect of resummation, as can be seen explicitly in Equation~\eqref{eq:gamma}, is to rescale the argument of the strong coupling to a softer scale $\as (Q^2/N)$. As we are going to show, this replacement corresponds to the introduction of a branch cut in the positive real $N$-axis, which makes it impossible to perform the inverse Mellin integral. For this reason, $\gamma(\as(Q^2),N)$ cannot be transformed back in physical space. Let's see why and where the branch cut occurs in the simpler case of LL accuracy: $\gamma$ has the form
\be 
\gamma_{\scriptstyle \rm LL}(\as(Q^2),N)=g_1 \int_1^N\frac{dn}{n}\,\as(Q^2/n)
=-\frac{g_1}{\beta_0}
\ln\left(1+\beta_0\as(Q^2)\ln\frac{1}{N}\right).
\label{gammaLL}
\ee
To be consistent, remember that we need to use the  the leading-log expression of $\as$:
\begin{equation}
\alpha_s(\mu^2)=\frac{\alpha_s(Q^2)}{1+\beta_0 \alpha_s(Q^2)\log \frac{\mu^2}{Q^2}}.
\label{eq:rgeLL}
\end{equation}
If we take $Q^2/n$ as the argument of $\as$, wee see that the denominator of Eq.~\eqref{eq:rgeLL} becomes
\begin{equation}
1- \beta_0 \as (Q^2) \log n.
\end{equation}
The singularity starts when this denominator vanishes, namely $\gamma_{\scriptstyle \rm LL}(\as(Q^2),N)$ has a branch cut on the real
positive axis for values of $N$ that satisfy
\be
\label{eq:nbranch}
N\geq N_L\equiv e^\frac{1}{\beta_0\as(Q^2)}.
\ee

From the presence of the branch cut follows that the inverse Mellin transform of the physical anomalous dimension does not exist. However, one may  consider the inverse Mellin transform of each term of
the expansion of $\gamma_{\scriptstyle \rm LL}(\as(Q^2),N)$ in powers of $\as(Q^2)$.
This would be
\be
\label{eq:Pasympt}
P_{\scriptstyle \rm LL}(\as(Q^2),x)=-\lim_{K\to\infty}\frac{g_1}{\beta_0}
\sum_{k=1}^K \frac{(-1)^{k+1}}{k}\beta_0^k \as^k(Q^2)\,
\frac{1}{2\pi i}\int_{\bar N-i\infty}^{\bar N+i\infty}dN\,
x^{-N}\,\ln^k\frac{1}{N};\qquad \bar N>0.
\ee
In this case, each term of the series is a well-defined inverse Mellin transform,
but the series as a whole is divergent, so  the limit $K\to\infty$ is not defined.  If the series were convergent,
one would be allowed to interchange the sum and the integral in eq.~(\ref{eq:Pasympt}), but then the sum
\be
\sum_{k=1}^\infty \frac{(-1)^{k+1}}{k}\beta_0^k \as^k(Q^2)\,
\ln^k\frac{1}{N}
\ee
would only be convergent when
\be
\abs{\beta_0\as(Q^2)\ln\frac{1}{N}}<1.
\ee

There are several ways to deal with the Landau pole divergence. Here we will discuss the so-called \emph{Borel prescription}. Another way is what is usually referred to as the \emph{minimal prescription}~\cite{minimalprescription}, which we will not cover in this thesis.
\section{Borel prescription}\label{sec:borelprescription}
 In order to understand the Borel prescription method, let us work with the divergent perturbative series
eq.~(\ref{eq:Pasympt}). Following step by step Ref.~\cite{borelprescription}, we compute the Mellin inversion integral:
\bea
\frac{1}{2\pi i}\int_{\bar N-i\infty}^{\bar N+i\infty}dN\,
x^{-N}\,L^k
&=&\left.\frac{d^k}{d\eta^k}\,
\frac{1}{2\pi i}\int_{\bar N-i\infty}^{\bar N+i\infty}dN\,
x^{-N}\,N^{-\eta}\right|_{\eta=0}\nonumber\\
&=&\frac{d^k}{d\eta^k}\,
\frac{1}{\Gamma(\eta)}\left[\ln^{\eta-1}\frac{1}{x}\right]_+
\Bigg|_{\eta=0}\nonumber +\delta(1-x)\,
\label{eq:Bconv1}
\eea
where we have used the identity 
\be
\int_0^1dx\,x^{N-1}\,
\left[ \ln^{\eta-1}\frac{1}{x}\right]_+=\Gamma(\eta)(N^{-\eta}-1).
\ee
Hence we obtain
\bea
P_{\scriptstyle \rm LL}(\as(Q^2),x)&=&\frac{g_1}{\beta_0}
\sum_{k=0}^K \frac{[-\beta_0 \as(Q^2)]^{k+1}}{k+1}\,
\Bigg\{\sum_{n=0}^{k+1}
\left(\begin{array}{c}k+1\\n\end{array}\right)
\left(\frac{d^n}{d\eta^n}\,\frac{1}{\Gamma(\eta)}\right)
\frac{d^{k+1-n}}{d\eta^{k+1-n}}
\left[\ln^{\eta-1}\frac{1}{x}\right]_+\Bigg|_{\eta=0}
\nonumber\\
&&
\qquad+\delta(1-x)\Bigg\} =
\nonumber
\eea
\bea
&=&\frac{g_1}{\beta_0}
\sum_{k=0}^K \frac{[-\beta_0 \as(Q^2)]^{k+1}}{k+1}
\Bigg\{\left[\frac{1}{\ln\frac{1}{x}}\sum_{n=1}^{k+1}
\left(\begin{array}{c}k+1\\n\end{array}\right)
 n \Delta^{(n-1)}(1) \left( \ln\ln\frac{1}{x}\right)^{k+1-n}\right]_+
\nonumber\\
&&\qquad+\delta(1-x)\Bigg\} =\nonumber \\
&=&\frac{g_1}{\beta_0}
\sum_{n=0}^K[-\beta_0 \as(Q^2)]^{n+1}\,
\Bigg\{
\frac{\Delta^{(n)}(1)}{n!}
\left[\frac{1}{\ln\frac{1}{x}}\sum_{k=n}^K
\frac{k!}{(k-n)!}\,
[-\beta_0 \as(Q^2)\ln\ln\frac{1}{x}]^{k-n}\right]_+
\nonumber\\
&&
+\frac{1}{n+1}\,\delta(1-x)\Bigg\} =
\nonumber\\
&=&
\frac{g_1}{\beta_0}
\sum_{n=0}^K\Bigg\{\Delta^{(n)}(1)\,
\left[\frac{1}{\ln\frac{1}{x}}
[-\beta_0\as(Q^2\ln\frac{1}{x})]^{n+1}\right]_+
+\frac{[-\beta_0 \as(Q^2)]^{n+1}}{n+1}\,\delta(1-x)\Bigg\}
\nonumber\\
&&+O(\as^{K+1}).
\label{eq:Pasymptsum}
\eea
Here we have defined $\Delta(z)\equiv1/\Gamma(z)$, and we
have used the identity
$\Delta^{(k)}(0)=k\Delta^{(k-1)}(1)$.
In the limit $K\to\infty$ the terms of order $\as^{K+1}$ can be
neglected, but we have already seen that the series is divergent.

In the large $x$ limit and up to leading logarithmic
accuracy we can rewrite
eq.~(\ref{eq:Pasymptsum}) as 
\be
P_{\scriptstyle \rm LL}(\as(Q^2),x)=\frac{g_1}{\beta_0}
\left[\frac{R(\as(Q^2),x)}{1-x}\right]_+
\label{lxdiv}
\ee
where
\be
R(\as(Q^2),x)=\lim_{K\to\infty}
\sum_{n=0}^K\Delta^{(n)}(1)\,[-\beta_0 \as(Q^2(1-x))]^{n+1}.
\label{eq:PasymptLL}
\ee

The idea of Ref.~\cite{borelprescription} is to sum the divergent series using the Borel method that we introduced in Section~\ref{sec:Borel}.   Namely, we take the Borel
transform of the divergent series (\ref{eq:PasymptLL}) with respect to
$\beta_0 \as(Q^2(1-x))$:
\be
\hat R(w,x)=-\sum_{j=0}^\infty\frac{\Delta^{(j)}(1)}{j!}
\,(-w)^j=-\frac{1}{\Gamma(1-w)}.
\label{eq:PwxLL}
\ee
To obtain the sum of the series, we have to compute the Borel integral
\be
R_{\scriptstyle \rm B}(\as(Q^2),x)=-\int_0^{+\infty}dw\,e^{-\frac{w}{\beta_0\as(Q^2(1-x))}}\,
\frac{1}{\Gamma(1-w)}.
\label{eq:invB}
\ee
Unfortunately, the integrand  diverges as $w\to\infty$. In particular, the
reflection formula
\be
\frac{1}{\Gamma(1-w)}=\frac{1}{\pi} \Gamma(w)\sin(\pi w)
\label{eq:ref}
\ee
implies that $\Delta(1-w)$ has a factorial oscillating behaviour when $w\to\infty$. 

We already saw in Section~\ref{sec:borelsummationexamples} how to deal with singularities that are along the path of
integration in the Borel inversion integral. In our case, the singularity is
at $w\to\infty$, hence we simply introduce an upper cutoff $C$ to the
integral. The divergent result eq.~(\ref{eq:invB}) is then replaced by
\be
R_{\scriptstyle \rm B}(\as(Q^2),x,C)=
-\int_0^C d w\,
\frac{1}{\Gamma(1-w)}\,
e^{-\frac{w}{\beta_0\as(Q^2(1-x))}},
\label{eq:pbreg}
\ee
which is  convergent for all $C$ and well defined for all $x$. Indeed, we can expand the integrand using
eq.~(\ref{eq:PwxLL}) and integrate 
term by term. The result is a convergent series, but, if  $C\to \infty$, then $\displaystyle \lim_{C\to\infty}f_k=1$ and we obtain once again
the original divergent series.

\section{Soft-gluon approximation}
\label{sec:softgluonapprox}
We have seen how the expansion of the resummed cross section in $z$ space is a divergent series. In some cases, at low perturbative orders the terms obtained by expansion of the resummed result turn out to provide a good approximation to the full result, even far from threshold. In the case of Higgs production, this has been known for long. Moreover, in Ref.~\cite{higgsn3loapprox} it was shown how the information from soft resummation can be combined with that from high-energy resummation to construct an approximation
\be \label{eq:Capprox}
C_\text{approx}(N,\as)= C_\text{soft}(N,\as)+C_\text{h.e.}(N,\as).
\ee

For the purpose of our thesis, we will only be interested in the soft part of this approximation. To see how we can use the resummed result to predict the full fixed-order computation, let us follow Ref.~\cite{higgsn3loapprox}. We first compare the resummed coefficient function in $N$ space with the exact expression $C^{(1)}(z)$, which is given by
\be \label{nlocf}
C^{(1)}(z) = 4A_g(z)\, \mathcal{D}_1(z) + d\, \delta(1-z) 
       -2A_g(z) \frac{\ln z}{1-z} + \mathcal{R}_{gg}(z),
\ee
where
\be
\mathcal{D}_k(z) \equiv \plus{\frac{\ln^k(1-z)}{1-z}} \label{eq:Dk}
\ee
and
\be
A_g(z) \equiv \frac{C_A}{\pi} \frac{1-2z+3z^2-2z^3+z^4}{z}.\label{eq:Ag_def}
\ee
Here, $d$ and $\mathcal{R}_{gg}(z)$ are functions of the dimensionless ratio $m_H/m_t$. The function $\mathcal{R}_{gg}(z)$ is regular in the limit $N\to\infty$, and therefore we won't need to worry about it.

The counterpart of this comparison is the expansion of Eq.~\eqref{eq:cresg0expS} to $\mathcal{O}(\as)$ at NLL logarithmic accuracy:
\begin{align}
\label{NLL1}
C_{\rm res}(N,\as)&=1+\as C^{(1)}_{\rm res}(N)+\mathcal{O}(\as^2),
\\
C^{(1)}_{\rm res}(N)&=g_{1,2}\ln^2N+g_{2,1}\ln N+g_{0,1},
\label{eq:C1res}
\end{align}
with
\be
g_{1,2} = \frac{2C_A}{\pi},\qquad
g_{2,1} = \frac{4C_A}{\pi} \gamma_{\scriptscriptstyle E}.
\ee
Now, if we compute the inverse
Mellin transform of Eq.~\eqref{eq:C1res}, we get
\begin{align}
C^{(1)}_{\rm res}(z,\as)
&= g_{0,1}\delta(1-z)+2g_{1,2} \mathcal{D}^{\log}_1(z) + \left(2\gamma_{\scriptscriptstyle E} g_{1,2}-g_{2,1}\right)\mathcal{D}^{\log}_0(z),
\nonumber\\
&= g_{0,1}\delta(1-z)+\frac{4C_A}\pi \mathcal{D}^{\log}_1(z),
\label{NLLz}
\end{align}
where
\be\label{eq:DkMP_def}
\mathcal{D}^{\log}_k(z) \equiv \plus{\frac{\ln^k\ln\frac1z}{\ln\frac1z}}.
\ee
If we now compare \eqref{eq:DkMP_def} to the soft contribution to the exact coefficient function Eq.~\eqref{eq:Dk}, we notice that they are different. The reason of this difference can be explained by noticing that singular terms as $z \to 1$ appear in the integral of the real emission diagrams over the gluon transverse momentum.

\begin{figure}[t]
\centering
\includegraphics[width=0.8\textwidth]
{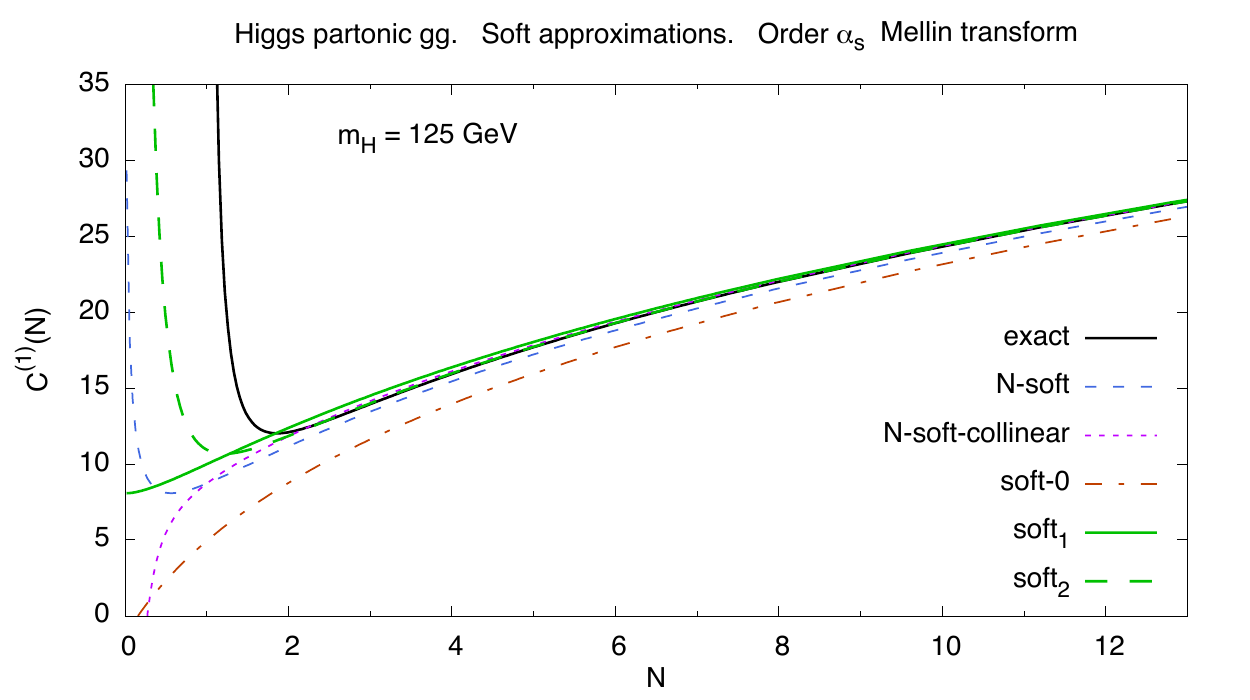} \\
\includegraphics[width=0.8\textwidth]
{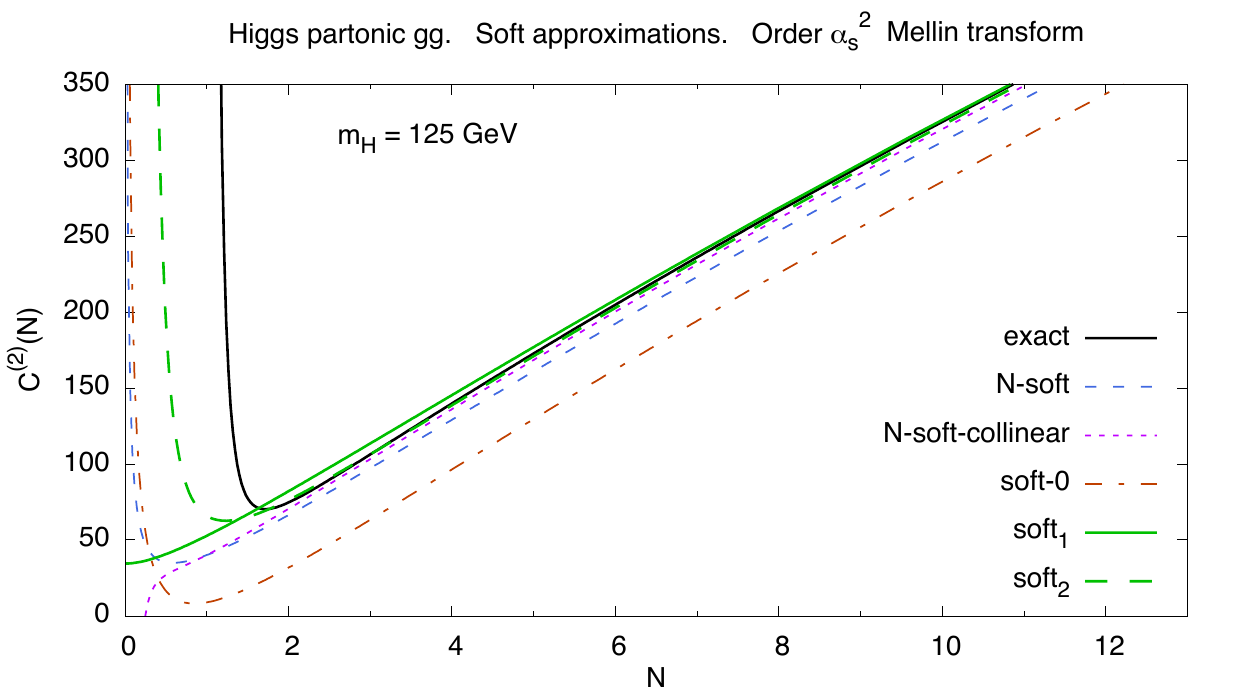}
\caption{Plot of the partonic coefficient functions $C^{(1)}(N)$ and $C^{(2)}(N)$
   for $m_H=125$~GeV. The black line represents the exact result, while the soft approximation is presented in various forms: the two preferred approximation of Ref.~\cite{higgsn3loapprox} are called soft$_1$ and soft$_2$, the so-called $N$-soft approximation (the one based on $\mathcal{D}^{\log}(N)$ as in Eq.~\eqref{eq:DkMP_def}), a collinear-improved $N$-soft approximation worked out by Ref.~\cite{softcollinear} and the soft-$0$ approximation which we didn't discuss.  Figure taken by~\cite{higgsn3loapprox}.}
\label{fig:soft_comparison}
\end{figure}

The approximation worked out by Ref.~\cite{higgsn3loapprox} is different from~Eq.\eqref{NLLz} by subleading terms. The validity of the approximation can be tested by comparing the soft approximation to the full result. Results as functions of the Mellin moment $N$ are shown in Fig.~\ref{fig:soft_comparison}, for the first two perturbative orders of the coefficient function.

\chapter{Renormalons}
\thispagestyle{empty}
\label{ch:renormalons}
Another source of divergence in QFT is known as renormalons. The ultimate origin of the divergence caused by renormalons can be found in the large momentum and small momentum behaviour of certain classes of corrections that can be inspected at all orders in perturbation theory.  Giving a definition to the concept of renormalon is not easy, so we will start with a classical example, the bubble diagram chain, in order to introduce the subject. A complete and exhaustive review on renormalons can be found in~\cite{beneke}. After introducing the issue, we will show that there is a deep connection between the renormalons and the complex plane where the Borel transform (that we saw in Section~\ref{sec:Borel}) is defined. Finally, we will explore the Borel plane and list all the known singularities that need to be taken into account when performing Borel summation.
\section{The bubble diagram chain}
We will now introduce the idea of renormalon divergence Ref.~\cite{beneke}. We will start with the computation of the bubble diagram chain and then we will give it an interpretation in terms of renormalons.

\begin{figure}
\centering
\includegraphics[width=\textwidth]{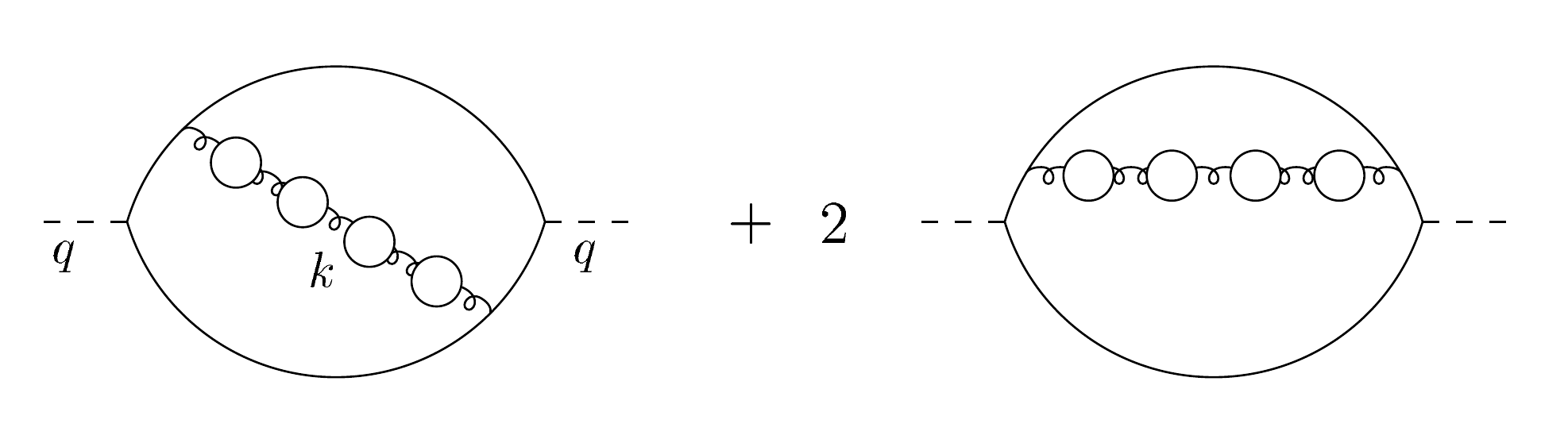}
\caption{The set of bubble diagrams for the 
Adler function consists of all diagrams with any number of fermion 
loops inserted into a single gluon line. Figure taken from Ref.~\cite{beneke}.}
\label{fig:bubblechain}
\end{figure}

In particular, we will consider the case of the correlation function of two currents of massless quarks  $j_\mu=
\bar{q}\gamma_\mu q$
\begin{equation}
\label{eq:currentcorr}
(-i)\int\,d^4x\,e^{-i q x}\,\langle 0|T\,(j_\mu(x) j_\nu(0))|0\rangle 
= \left(q_\mu q_\nu-q^2 g_{\mu\nu}\right)\,\Pi(Q^2)
\end{equation}
where, as usual, $Q^2=-q^2$. We will consider a certain class of corrections to the Adler function, which is defined as
\begin{equation}
\label{eq:adlerdef}
D(Q^2)=4 \pi^2\,\frac{d\Pi(Q^2)}{dQ^2}.
\end{equation}
The class of corrections that we would like to compute is represented in Fig.~\ref{fig:bubblechain}. Keep in mind that this is just an illustrative example: they don't represent the only contribution to the renormalon divergence. However, historically, this is how renormalons were first introduced in the original works by Refs.~\cite{grossneveu,lautrup,thooft}. 
To compute the Adler function we need to insert multiple times the renormalized fermion loop:
\begin{equation}
\label{eq:simpleloop}
\beta_{0f}\alpha_s\left[\ln(-k^2/\mu^2) + C\right],
\end{equation}
where $C$ is constant that depend on the renormalization sheme ($C=-5/3$ in the $\overline{\rm MS}$ scheme) and $\beta_{0f}$ is
the fermion contribution to $\beta_0$
\begin{equation}
\beta_{0f} = - \frac{N_f}{6 \pi}.
\end{equation}
 For the time being, we are only considering fermion bubbles, so we only need the fermion contribution to the $\beta$ function. We will see what happens when we include gluon and ghost terms in Section~\ref{sec:gluonandghost}.

When we want to include $n \to \oo$ fermion loops, we integrate over the loop momentum 
of the big external fermion loop and over the angles of 
$k$. Let us define $\hat{k}^2=-k^2/Q^2$, we get
\begin{equation}
\label{eq:basint}
D = \sum_{n=0}^\infty \,\alpha_s\int\limits_0^\infty \,\frac{d\hat{k}^2}
{\hat{k}^2}\,F(\hat{k}^2)\,\left[\beta_{0f}\alpha_s\ln\left(
\hat{k}^2\frac{Q^2e^{-\frac53}}{\mu^2}\right)\right]^n.
\end{equation}
The function $F$ is rather complicated in its exact form, but we can just consider the  $n\gg 1$ approximation. due to large logarithmic enhancements, the dominant contributions to the integral come from very large or very low values of $k$ ($k\gg Q$ and $k\ll Q$). For this reason, we are only interested in the  
small-$\hat{k}$ and large-$\hat{k}$ behaviour of $F$:
\begin{align}
\label{eq:smallk}
F(\hat{k}^2) &= \frac{3 C_F}{2\pi}\,\hat{k}^4 + {\cal O}(\hat{k}^6 \ln \hat{k}^2) & \hat{k} \to 0,\\
\label{eq:largek}
F(\hat{k}^2) &= \frac{C_F}{3\pi}\,\frac{1}{\hat{k}^2} \left(\ln\hat{k}^2+\frac{5}{6}\right) +
{\cal O}\!\left(\frac{\ln\hat{k}^2}{\hat{k}^4}\right) &\hat{k} \to \infty.
\end{align}
It is worth while to notice that we expect a power-like approach to zero in both cases since the Adler function is UV and IR finite. If we split the integral \eqref{eq:basint} at $\hat{k}^2=\mu^2/(Q^2 e^{-5/3})$ and we
insert \eqref{eq:smallk} for the small-$\hat{k}^2$ interval and 
\eqref{eq:largek} for the large-$\hat{k}^2$ interval, we obtain
\begin{equation}
\label{eq:adlerapprox}
D \sim \frac{C_F}{\pi} \sum_{n=0}^\infty \alpha_s^{n+1} 
\left[\frac{3}{4}\left(\frac{Q^2}{\mu^2} e^{\frac53}\right)^{-2}
\left(\frac{\beta_{0f}}{2}\right)^n n! + 
\frac{1}{3}\,\frac{Q^2}{\mu^2} e^{\frac53}\,
(-\beta_{0f})^n\,n!\,\left(n+\frac{11}{6}\right)\right],
\end{equation}
where the first term comes from the small $\hat{k}$ region and is accurate up to 
relative corrections of order $n\,(2/3)^n$, while the second term comes from the
large $\hat{k}$ region and is accurate up to $(1/2)^n$.
This is the reason why the factorial divergences of the 
two series are called respectively {\em infrared (IR) renormalon} and 
{\em ultraviolet (UV) renormalon}. The etymology of the word 
``renormalon'' goes back to Ref.~\cite{thooft}. There, the word renormalon was chosen as an analogy to  the only other known (at that time) source of divergence: the instanton divergence. This divergent 
behaviour was then new and typical of  
renormalizable field theories.

Equation~\eqref{eq:adlerapprox} exhibits the same factorial divergence that we encountered in the examples of Section~\ref{sec:borelsummationexamples}. The first order Borel transform reads
\begin{align}
B[D](t) &= \frac{3 C_F}{2 \pi}
\left(\frac{Q^2}{\mu^2} e^{-\frac53}\right)^{-2}\frac{1}{2-\beta_{0f} t} &\mbox{(first IR renormalon)}  \nonumber\\
&+ \,\frac{C_F}{3\pi} \,\frac{Q^2}{\mu^2} e^{-\frac53}\,\left[ \frac{1}{(1+\beta_{0f} t)^2}+\frac{5}{6}\frac{1}{(1+\beta_{0f} t)}\right] &\mbox{(first UV renormalon)}, \label{eq:firstpolesD}
\end{align}
The corresponding singularities in 
the Borel plane lie at $t= 2/\beta_{0f}$ (IR renormalon) and 
$t=-1/\beta_{0f}$ (UV renormalon). 
  Eq.~\eqref{eq:firstpolesD}
only gives us the singularities that are
close to the origin. It can be shown that the exact Borel transform 
of the set of diagrams of Fig.~\ref{fig:bubblechain} is made of an infinite sequence of IR (UV) renormalon poles at negative 
(positive) integer multiples of $m \beta_{0f} t$, apart from $m=1$. Therefore, we define the term \emph{renormalon} more generally as a singularity 
of the Borel transform, which is ultimately related to the large or the small momentum 
behaviour of the loop. As we have specified above, the set of bubble graphs only provides 
some of such singularities, but not all of them. 

Now that we have introduced the subject of renormalon divergence with an explicit example, let us make a few observations, which will be useful for the application of renormalons to determine theoretical uncertainties in QCD.
\subsection{Adding gluon and ghost terms}
\label{sec:gluonandghost}
So far we have only considered one loop corrections due to fermion bubble graphs, and therefore we have consistently used the fermion 
contribution $\beta_{0f}$ to the $\beta$-function. The next step is to 
add the gluon and ghost bubbles, but that introduces a complication, since the result 
is gauge-dependent. 

What happens is that the effect of substituting $\beta_{0f}$ with $\beta_0$ flips the location of the renormalon singularities. The fermion contribution $\beta_{0f}$ is negative, but $\beta_0$ is positive, and therefore UV renormalons are now located in the negative real axis, while IR renormalons are in the positive real axis of the Borel plane. Moreover, UV renormalons give origin now to a sign-alternating divergence, while IR renormalons
 introduce an ambiguity in the Borel integral as we saw in Section~\ref{sec:borelsummationexamples}. To remove this ambiguity, one needs to add non-perturbative corrections. The same situation occurs in QED with UV renormalons. In both cases, as one could easily expect, non-perturbative corrections are needed when the coupling becomes large (infrared in QCD and ultraviolet in QED).

At this stage, there is no evident reason to suppose that we can extrapolate to the full non-abelian $\beta_0$. However, it is shown by~\cite{beneke} that the 
substitution of $\beta_{0f}$ by $\beta_0$ can be fully justified. For further simplification, since the poles of the Borel transform are located at multiples of $\beta_0 t$, we will use the definition
\begin{equation}
u = \beta_0 t.
\end{equation}
This way, IR renormalons are located at $u = n$, while UV renormalons are located at $u= - n$, where $n$ is a positive integer $\neq 0$.
\subsection{Power corrections}
Let us now consider a generic perturbative series where we have already collected an overall $\as$ factor
\begin{equation}
R = \sum_{k=0}^\oo r_k \, \alpha_s^{k+1}.
\end{equation}
This way the Borel integral has a slightly different form than in section~\ref{sec:Borel}:
\begin{equation}
\tilde{R} = \int_0^{\infty} \d t \, e^{-t/\as} \, B[R] (t).
\label{eq:borelintegral}
\end{equation}
If the Borel transform $B[R](t)$ presents the typical IR renormalon behaviour, 
\begin{equation}
B[R] (t) = \frac{K}{1- a \beta_0 t},
\end{equation}
where $1/a$ is a positive integer and the radius of convergence is $|t|<1/|a \beta_0|$, then we have an ambiguity in the determination of the integral. $\tilde R$ acquires an imaginary part, which is the residue at the pole of the Borel integral:
\begin{equation}
\text{Im} (\tilde R) = \pm \pi  \text{Res}\left( \frac{1}{a \beta_0} \right) = \pm \pi  \frac{K}{a \beta_0} \exp \left(- \frac{1}{a \beta_0 \as} \right).
\end{equation}
Now we might be interested in the scaling behaviour of this ambiguity, which in some sense represents the difference between the exact value of the series and the asymptotic value given by the Borel integral. To do so, let us recall the definition of the running coupling constant in terms of $\Lambda_\text{QCD}$:
\bea\label{eq:alpha_Lambda}
\alpha_s(\mu^2) = \frac{1}{\beta_0 \log \displaystyle \frac{\mu^2}{\Lambda_{\text{QCD}}^2}}.
\eea
If the Borel transform has a singularity at $u=1/a$, that yields an ambiguity in the definition of the 
Adler function which scales as 
\begin{equation}
\label{eq:powercorr}
\delta D(Q^2) \propto \exp \left(- \frac{1}{a \beta_0\alpha_s(Q^2)}\right) = \left(\frac{Q}{\Lambda_\text{QCD}}\right)^{-\frac 2a} = \left(\frac{\Lambda_\text{QCD}}{Q}\right)^{\frac 2a}.
\end{equation}
Recall that the first IR renormalon pole in \eqref{eq:firstpolesD} corresponds to $a = 1/2$, and therefore to a leading power correction proportional to
\begin{equation}
\delta D(Q^2) \sim 
\left(\frac{\Lambda_\text{QCD}}{Q}\right)^{4} 
\end{equation}
\section{The Borel plane}
\begin{figure}
\centering
\includegraphics[width=\textwidth]{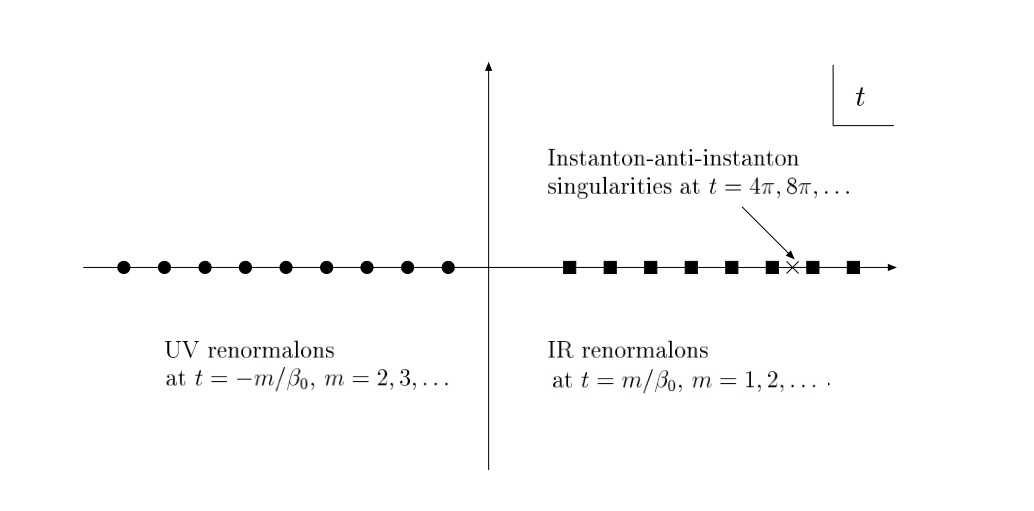}
\caption{Singularities in the Borel plane of 
the Adler function in QCD. The singular points are shown, but not the cuts attached to each 
of them. Figure taken from Ref.~\cite{beneke}.
\label{fig:borelplane}}
\end{figure}
We have seen how a renormalon is nothing but a pole on the real axis of the Borel inverse integral. For what concerns the Adler function, the Borel plane is shown in Figure~\ref{fig:borelplane}. Of course, there might be a whole new set of singularities that we don't know, but among all the known singularities we distinguish three sets:

UV renormalons are poles located in the negative real axis at $t=-m/\beta_0$, with positive 
integer $m$. The leading term of the power correction associated with UV renormalon should be, following  
\eqref{eq:firstpolesD}, of order $\Lambda_\text{QCD}^2/Q^2$, 
using \eqref{eq:powercorr}. Actually, it can be shown that its precise form is
\begin{equation}
\label{eq:uvpowercorr}
\delta D_{\rm UV} \sim \frac{Q^2\Lambda^2}{\mu^4}\,\times\,
\mbox{logarithms}.
\end{equation}
However, the Borel inverse integral only involves positive values of $u$, so UV renormalons produce no ambiguity in the Borel integral and a sign-alternating factorial divergence. Therefore, no extra terms should 
be added to the perturbative expansion. UV renormalons depend on the theory, but are
process independent.

IR renormalons can be found in the positive real axis, at $t=m/\beta_0$, with $m=2,3,\ldots$. The leading 
term associated with the first 
IR renormalon is of order $(\Lambda/Q)^4$.  Unlike UV renormalons, IR renormalons are process-dependent.
\subsection{The instanton divergence}
\label{sec:instanton}
In Figure~\ref{fig:borelplane} we can observe a third set of singularities, which leads to a factorial divergence in the perturbative series: the so-called \emph{instantons}~\cite{lipatov}. Instantons are classical solutions to equations of motion with a finite, non-zero action. Ref.~\cite{bogofate} shows how configurations of $n$ instantons 
and $n$ anti-instantons give rise to poles in the positive real axis in the Borel plane located at $t=4\pi n$. Instanton divergence can be related to  the factorially increasing number of diagrams in perturbation theory. Thanks to the semi-classical origin of instantons, the residues at the poles in the Borel plane can be calculated. This calculation is worked out by \cite{balitsky} for the Adler function. 
However, in QCD instanton poles are located at large $t$, where $t$ is the variable of the Borel transform. We shall see in Section~\ref{sec:divergencepoint} that this means that their effect is by all means negligible for what concerns large orders in perturbation theory in QCD. Moreover, recalling Eq.~\eqref{eq:powercorr}, it is easy to see that  they 
do not represent a dominant source of power corrections. For a complete review on the subject of instanton, see~\cite{LGZJ}.
 
\chapter{Theoretical uncertainties}
\label{ch:theorunc}
\thispagestyle{empty}
Whenever performing an experiment, such as at the LHC, we compare measurements to theoretical calculations and try to find out if they match or not. However, in QFT the theory is perturbative, which means that theoretical predictions are perturbative series of which only the first terms are known. In this case, a full control of the uncertainty of these predictions becomes of paramount importance, as both the experiment and the theory need to be provided with a degree of uncertainty in order to determine their agreement. In QCD the issue of theoretical accuracy is pressing, due to the large size of the coupling $\alpha_s$ and therefore its slow perturbative convergence.

There must exist, therefore, a definition of  \emph{theoretical uncertainty} on any calculation of any observable at the LHC. Following Ref.~\cite{higgsworkinggroup}, we first make a distinction between parametric uncertainties (PU),
related to the value of input parameters, and actual theoretical uncertainties (THU),
related to our lack of knowledge about higher orders in perturbation theory. There is no way of eliminating completely parametric uncertainties, even though  they can be reduced when
more precise experiments produce improved results. Theoretical uncertainties, however, might be ideally eliminated if an all-order computation were available.

Another difference between PU and THU is that PU are distributed according to a 
known (usually Gaussian) distribution while THU are arguably distributed according to a flat distribution, even if the statistical interpretation of THU is less clear. PU have been studied in detail during the past several years, and we will not talk about them in this thesis. Among the models that give an estimate to the THU, we will consider three: the scale variation method, the Cacciari-Houdeau method and the David-Passarino method.

The scale variation is the conventional and most widely used method to determine theoretical uncertainties. It is based on the following idea: in the full theory there should not be any scale dependence
and order by order in perturbation theory we should be able to see the
asymptotic limit. Therefore, variation of the scale (or scales) is a pragmatic
way of understanding how far we are from controlling the theory.

The Cacciari-Houdeau method, introduced in Ref.~\cite{cacciarihoudeau}, is based on a Bayesian model that, given a certain distribution for the coefficient of the series, allows one to characterise a perturbative theoretical uncertainty in terms of a credibility interval for the remainder of the series.

The David-Passarino method~\cite{davidpassarino}, instead, uses the concept of sequence transformation to improve the convergence of the series. Using sequence transformation, slowly convergent series can be transformed into series that have better numerical properties. Even if a sequence transformation hardly ever sums a series exactly, it usually predicts some of the unknown terms of the sequence.

In order to fix the notation, from now on we will define the partial sum of a perturbative cross section up to the $k$ order as
\be
\sigma_k \equiv \sum_{n=l}^k c_n \alpha_s^n,
\label{eq:sigmapartialsum}
\ee
and the remainder of the series as
\be
\Delta_k \equiv \sum_{n=k+1}^\infty c_n \alpha_s^n\,.
\label{eq:remainder}
\ee

\section{Scale variation}
\label{sec:scalevariation}
In QCD, there is no obvious optimal choice for the renormalization and factorization scale. In QED, we have a  physical subtraction point, $q^2 = 0$,
for photons with momentum transfer $q$, which is referred to as the Thomson limit. In the EW theory, once again, we have a 
 physical subtraction point(s):  the electromagnetic coupling is still fixed in the Thomson limit, while the weak mixing angle is related to the ratio of the $W$ and $Z$ boson masses. In other words, in the electromagnetic and EW theory our calculations do depend on $\mu_{R}$, but this dependence disappears once the Lagrangian parameters are replaced by data.

However, in QCD we have
no analogue of $G_{F}$. This means that if $s$ is the scale at which we study the process, our LO calculation will always contain logarithms like 
$\ln(s/\mu_{R})$. One should find a scale where some data is available (a subtraction point), but this is not yet possible in QCD. So the question arises of what is the best choice for $\mu_R$?

The general guideline~\cite{higgsworkinggroup} is to set $\mu_{R}$  exactly to the relevant scale $s$. Of course, this is straightforward in processes where only one scale is relevant, but in processes where multiple scales are relevant there will be additional logarithms of argument $s/s'$.  In this case, the convention is to choose the renormalization scale $\mu_{R}$ and the factorization scale $\mu_{F}$ process by process, in such 
a way to minimize the
effect of the new corrections when going to the next perturbative order. This choice of scale is sometimes called {\em dynamical scale} and it is the closest thing to a subtraction point in QCD.

The scale variation method consists in giving an estimate to the THU in terms of an interval centered around the dynamical scale, e.g.~$s/n < \mu_{R,F} < n\,s$. There is no specific choice for the value of $n$, but it should be fixed so as to include a plateau in the scale dependence.

When combining the renormalization scale and factorization scale, once again we should try to minimize the effect of new corrections. The possible choices for this combination are a diagonal scan, a diagonal scan with anti-diagonal corrections or a two dimensional scan with $1/n < \mu_{R}/\mu_{F} < n$. 
\section{Cacciari-Houdeau model}
\label{sec:cacciarihoudeau}
The main shortcoming of the scale variation is that it does not provide the degree of belief of the resulting uncertainty bands. In other words, it does not associate a numerical value to the probability that the uncertainty band contains the true sum of the series. Because of that, it becomes difficult to combine THU with the above mentioned PU.
\\
The Cacciari-Houdeau Method \cite{cacciarihoudeau} provides theoretical uncertainties with a well defined credibility measure and computes explicitly the degree of belief of a given interval.

Of course, this degree of belief has nothing to do with the concept of frequentist probability, but it needs to be intended in terms of the \emph{Bayesian probability}. While frequentist probability is linked to a large number of realizations of an experiment, Bayesian probability deals with the mathematical treatment of the ``trueness'' of a statement. The variables that allow us to treat the frequentist probability are called \emph{random variables}, while we can call \emph{uncertain variables} the ones that appear in a credibility distribution.
\subsection{Model overview}
\label{sec:hypotheses}
Given a series like in Eq.~\eqref{eq:sigmapartialsum}, the model gives a form to the distribution of the coefficients $c_0, c_1, \dots$.
The three main hypotheses of the CH models are
\begin{enumerate}
\item The residual density for the unknown coefficients $c_n$, given a known upper bound $\bar c$, is a flat distribution.
\item The upper bound $\bar c$ is the only parameter that contains all the information.
\item All values for the parameter $\bar c$ are equally probable.
\end{enumerate}
Once we have these three hypotheses, we have defined the credibility measure over the space of uncertain variables $\{ c_0, c_1, \dots\}$. 
\\

Let us consider now a perturbative series whose first order is not $n=0$ but $n=l$. Using the three hypotheses, and defining $\bar c_{(k)} \equiv \max (|c_l|, \dots, |c_k|)$, the conditional density for the remainder $\Delta_k$ can be obtained (see \cite{cacciarihoudeau} for a complete derivation):
\be
f(\Delta_k|c_l, \dots, c_k) = \int \left [ \delta \left (\Delta_k - \sum_{n=k+1}^\infty \alpha_s^n c_n \right ) \right ] f(c_{k+1}, c_{k+2}, \dots | c_l, \dots, c_k) \, \md c_{k+1} \, \md c_{k+2} \dots \,.
\ee
It is difficult to treat this expression analytically, but we can make the approximation that the whole remainder of the series is comparable to the first term of the remainder itself:

\be
|\Delta_k| \simeq \alpha_S^{k+1} |c_{k+1}| \,.
\ee
In this case the number of known coefficients is $n_c=k+1-l$, and we obtain
\be
\label{eq:conditionalresult}
f(\Delta_k|c_l, \dots, c_k) \simeq \left ( \frac{n_c}{n_c+1} \right ) \frac{1}{2 \alpha_s^{k+1} \bar c_{(k)}} \left\{
\begin{array}{cc}
1	&	\mbox{ if }	|\Delta_k|\leq \alpha_S^{k+1}\bar c_{(k)}\\[10pt]
\frac{1}{(|\Delta_k|/(\alpha_s^{k+1}\bar c_{(k)}))^{n_c+1}}	&	\mbox{ if }	|\Delta_k|>\alpha_s^{k+1}\bar c_{(k)}
\end{array} \right. . 
\ee
If we know  $f(\Delta_k |c_l,\dots,c_k)$, we can calculate any $p$-credible interval for $\Delta_k$, where $p$ is the Bayesian probability that the true remainder of the series will be inside the interval:
\be
p = \int_{-d_k^{(p)}}^{d_k^{(p)}}  f(\Delta_k |c_l,\dots,c_k) \md\Delta_k \,.
\ee
Finally, using a suitable approximation, it can be shown that
\be
\label{eq:cacciariresult}
d_k^{(p)}= \left\{
\begin{array}{cc}
\alpha_S^{k+1}\bar c_{(k)}\,\frac{n_c+1}{n_c}\,p                   &	\mbox{ if }	 p \le \frac{n_c}{n_c+1}\\[10pt]
\alpha_S^{k+1}\bar c_{(k)}\left[(n_c+1)(1-p)\right]^{-1/n_c} &	\mbox{ if }	 p > \frac{n_c}{n_c+1} 
\end{array}
\right. \,,
\ee
where $p \equiv P/100$ and $P$ is a number between 0 and 100.

Therefore, the CH model not only gives us the theoretical uncertainty interval for a generic perturbative series, but it also tells us what is the Bayesian confidence level of that interval. However, the first hypothesis is that the coefficient of the series $c_n$ should be of the same order of magnitude. This turns out to be not applicable to series like the Higgs cross section (see Table~\ref{tab:higgsxs}). The modified Cacciari-Houdeau approach, introduced in~\cite{cacciaribagnaschi}, aims to solve this issue.
\subsection{The modified Cacciari-Houdeau approach \texorpdfstring{($\overline{\mathrm{CH}}$)}{}}
\label{sec:chbarfac}
In many cases the coefficients $c_n$ are not of the same size, but instead they present a noticeable growth with the increasing perturbative order. Since the CH model relies on a specific form for the perturbative expansion, Eq.~\eqref{eq:sigmapartialsum}, results are not invariant under
a rescaling of the
expansion parameter from $\as$ to $\as/\lambda$. 
Ref.~\cite{cacciaribagnaschi} presents a slightly modified version of the CH model. In this modified model, 
denoted as $\overline{\mathrm{CH}}$, we first rewrite the perturbative expansion of eq.~(\ref{eq:sigmapartialsum}) in the form
\begin{equation}
  \label{eq:chbarexp}
  \sigma_k=\sum\limits_{n=l}^{k}\frac{\alpha_s^n}{\lambda^n} (n-1)!
  \frac{\lambda^n c_n}{(n-1)!}
  \equiv
  \sum\limits_{n=l}^{k}\left(\frac{\alpha_s}{\lambda}\right)^n (n-1)!\, b_n\, ,
\end{equation}
with
\be
b_n \equiv \frac{\lambda^n c_n}{(n-1)!} \, .
\ee
Then we use the coefficients $b_n$ as the $c_n$ coefficients in the CH model. The probability density becomes therefore
\be
f(\Delta_k|b_l,\dots,b_k)\simeq\left(\frac{n_c}{n_c+1}\right)
\frac{1}{2k!(\as/\lambda)^{k+1}\bar b_{(k)}}\left\{
\begin{array}{cc}
  1    & \mbox{ if }	|\Delta_k|\leq k!\left(\frac{\as}{\lambda}\right)^{k+1}\bar b_{(k)} \\[10pt]
  \left(\frac{k!(\as/\lambda)^{k+1}\bar b_{(k)}}{|\Delta_k|}\right)^{n_c+1} &
  \mbox{ if }	|\Delta_k|>k!\left(\frac{\as}{\lambda}\right)^{k+1}\bar b_{(k)}
\end{array}
\right. \, 
\ee
and the credibility interval
\begin{align}
  \label{eq:intervalCHbar}
  d_k^{(p)}& =\left\{
  \begin{array}{l l} 
    k! \left(\frac{\alpha_s}{\lambda}\right)^{k+1} \bar{b}_{(k)} \frac{n_c+1}{n_c} p\% & 
    \text{if} \qquad p\% \leq \frac{n_c}{n_c+1} \\\\
    k! \left(\frac{\as}{\lambda}\right)^{k+1} \bar{b}_{(k)} \left[(n_c+1)(1-p\%)\right]^{(-1/n_c)} & 
    \text{if} \qquad p\% > \frac{n_c}{n_c+1}   \\
  \end{array}\right.\qquad .
\end{align}

There are several ways to determine the optimal value for the rescaling factor $\lambda$.  Ref.~\cite{cacciaribagnaschi} determines $\lambda$ empirically by observing how the model succeeds in predicting theoretical uncertainties for observables   for which higher order perturbative computations are known. 
Another way of determining $\lambda$ is proposed in~\cite{doweneed}, where the best $\lambda$ is thought to be the one that makes all the expansion coefficients closest to the same size.
\section{David-Passarino model}
Reference~\cite{davidpassarino}  predicts higher orders using the concept of series acceleration. In this case they use of a collection of series transformations in order to improve the convergence of a series. Even if the original series may be (and in the QED/QCD case is) divergent, the helpful property of sequence transformations is that they provide a result that can be interpreted as the analytic extension of the sum of the series. Therefore, the difference between the all-order sum and the partial sum up to the last known order is what we call the theoretical uncertainty.

The hypotheses on which this model is based are related to the analyticity of the series. In fact, there are an infinite number of functions that have the same asymptotic expansion, therefore we need to assume that
\begin{enumerate}
\item The analyticity domain is sufficiently large
\item After a certain perturbative order, there is an upper bound on the remainder of the series.
\end{enumerate}

The general applicability of sequence transformations to QFT theories is widely discussed in Ref.~\cite{hypothesisDP}. We briefly review the most important sequence transformations that are used in the model.
\subsection{The Levin $\tau\,$-transform}
The first kind of sequence transformation that one can consider is the so-called  Levin $\tau\,$-transform.
If the partial sum has the form
\be
S_n = \sum_{i=0}^n\,\gamma_i\,z^i,
\label{eq:DPpartialsum}
\ee
we can define the $\tau\,$-transform as
\be
\tau^n_k(\beta) = \frac{\sum_{i=i_0}^k\,W^{\tau}\left( n,k,i,\beta\right)\,S_{n+i}}{\sum_{i=i_0}^k\,
W^{\tau}\left( n,k,i,\beta\right)},
\qquad \tau_k \equiv \tau^{0}_k \equiv \tau^0_k(0)
\label{Levintau}
\ee
where $i_0 = n-1$ and
\be
W^{\tau}(n,k,i,\beta) = (-1)^i\,
\binom{k}{i}
\frac{\left( \beta+n+i\right)_{k-1}}{\Delta S_{n+i-1}},
\ee
where $(z)_a = \Gamma(z+a)/\Gamma(z)$ is the Pochhammer symbol. $\Delta$ represents the forward-difference operator: $\Delta S_n = S_{n+1} - S_n$.

The key to estimate the first unknown coefficient is to Taylor expand $\tau_k$. Suppose that $S_1\,,\,\dots\,,\,S_k$ are known. One then computes
\be
\tau_k - S_k = {\bar{\gamma}}_{k+1}\,z^{k+1} + {\cal O}\left( z^{k+2}\right)
\ee
and ${\bar{\gamma}}_{k+1}$ is the prediction for $\gamma_{k+1}$.
If the number of known values $k$ is very small, then the prediction is not expected to be reliable. However, if we apply 
\be
\tau_2 - S_2 = (\gamma^2_2/\gamma_1)\,z^3 + {\cal O}\left( z^4\right)
\ee
to the Higgs series, we end up with a result that has the correct sign and order of magnitude.

The procedure that allows us to improve the convergence of the series is the following: suppose that we want to apply the Levin $\tau\,$-transform $\tau^0_k(\beta)$ to the Higgs series.
\begin{enumerate}

\item First use the first $3$ terms in \eqref{eq:DPpartialsum}, with $\gamma_3= \gamma^{c}_3(\mu =m_H)$, and derive $\bar{\gamma}_4$:
\be
\bar{\gamma}_4 = 3\,\frac{\gamma_3}{\gamma_1 \gamma_2}\Bigl[
 2\,\frac{\left( 5 + 2\,\beta\right)\,\gamma^2_2 - \left( 3 + \beta\right)\,\gamma_1\,\gamma_3}
{12 + 7\,\beta + \beta^2}
 + \gamma_1\,\gamma_3 - \gamma^2_2 \Bigr].
\ee
\item Compute $S_4$ assuming $\gamma_4 = {\bar{\gamma}}_4$.
\item Derive 
\be
\bar{\gamma}_5 = 
\frac{\vartheta\,\bar{\gamma}_4}{\gamma_1\,\gamma_2\,\gamma_3}\,
\left( 120 + 72\,\beta + 15\,\beta^2 + \beta^3\right)^{-1},
\ee
where
\bea
\vartheta &=& 4\,\gamma_2^2\,\gamma_3 \, \Bigl(6 + 11\,\beta + 6\,\beta^2 + \beta^3\Bigr)
     - 6\,\gamma_1\,\gamma_3^2 \, \Bigl(24 + 26\,\beta + 9\,\beta^2 + \beta^3\Bigr) \nonumber \\
     &+& 4\,\gamma_1\,\gamma_2\,\bar{\gamma}_4 \, \Bigl(60 + 470\,\beta + 12\,\beta^2 + \beta^3\Bigr).
\eea
\item Calculate $S_5$ assuming $\gamma_5 = {\bar{\gamma}}_5$.
\item Repeat steps 1 to 4 until $\tau_3,\,\dots,\tau_6$ are computed.
\item Compare the $S_3,\,\dots,S_6$ with the $\tau_3,\,\dots,\tau_6$.
\item Repeat again steps 1--6 for $\gamma_3 = \gamma^{c}_3 +
\Delta\gamma_3$ and $\gamma_3 = \gamma^{c}_3 -
\Delta\gamma_3$, always taken at $\mu=m_H$.

\end{enumerate}

\subsection{The Weniger $\delta\,$-transform}
A second transform that is considered in~\cite{davidpassarino} is the $\delta\,$-transform introduced by 
Weniger:
\be
\delta_k(\beta) = \frac{\sum_{i=0}^k\,W^{\delta}\left( k,i,\beta\right)\,S_{i}}{\sum_{i=0}^k\,
W^{\delta}\left( k,i,\beta\right)},
\qquad \delta_k \equiv \delta_k(1),
\label{Wenigerdelta}
\ee
where 
\be
W^{\delta}(k,i,\beta) = (-1)^i\,
\binom{k}{i}
\frac{\left( \beta+i\right)_{k-1}}{\left( \beta+k\right)_{k-1}}\,\frac{1}{\gamma_{i+1}\,z^{i+1}}.
\ee
Of course, we can follow the same exact recipe described above, and we get
\be
\bar{\gamma}_4= 
   \frac{\gamma_3}{3\,\gamma_1\,\gamma_2}\,\left( 4\,\gamma_1\,\gamma_3 - \gamma^2_2\right)
\ee
 and 
 \be
\bar{\gamma}_5 = 
 \frac{\bar{\gamma}_4}{10\,\gamma_1\,\gamma_2\,\gamma_3}\,\left(
 \gamma^2_2\,\gamma_3 - 9\,\gamma_1\,\gamma^2_3 + 18\,\gamma_1\,\gamma_2\,\bar{\gamma}_4 \right).
\ee

\chapter{Phenomenological applications}
\label{ch:results}
\thispagestyle{empty}
In this Chapter we turn our attention to possible phenomenological applications of what we have learned in the previous Chapters. In particular, our first concern will be to determine the perturbative order at which the partial sum start to diverge. We know that at some point the perturbative expansion of any observable in QCD will deviate from its real value, but we do not know when. If, for example, a series started to diverge at the fourth perturbative order, then we should start to worry that our theoretical predictions might be meaningless.

After studying the divergent behaviour of a series, one can give an estimate to its asymptotic value, via the Borel method illustrated in Section~\ref{sec:Borel}. Hence, the theoretical uncertainty on the perturbative expansion is defined as the difference between the asymptotic value and the truncated series.

Since we have seen three sources of divergence but the instanton impact is by all means negligible, we will first study the effect of the Landau pole divergence and renormalon divergence assuming that only one source at a time is dominant. Then we will combine both sources to obtain the most complete estimate for the theoretical uncertainty. The combination is important because we know that the soft approximation is dominant at low orders, but we also know that the renormalon divergence will occur sooner.

We will apply our models to two of the most relevant processes at the LHC: Higgs production and $t \bar t$ production. The former is known exactly up to N$^3$LO~\cite{anastasiou} and its soft approximation is known to work very well~\cite{higgsn3loapprox}. The highest PDF order available is NNLO, so one might ask if it is sensible to convolute the N$^3$LO partonic cross section with the NNLO parton distributions. The answer~\cite{doweneed} is yes, because the impact of theoretical uncertainties on PDFs is negligible for the Higgs cross section. For $t \bar t$ production, the perturbative series is known up to NNLO and the soft approximation has been studied in~\cite{topn3loapprox}.

Finally, once we have determined our theoretical uncertainty, we will have to check its consistency. We do so first by pretending not to know the N$^3$LO and trying to predict it.
Then, we compare our method with the other available methods mentioned in Chapter~\ref{ch:theorunc}.
\section{Landau pole divergence only}
The first case we consider is a model where only the Landau pole divergence is present. In this scenario, the all-order behaviour of the cross section is dictated by the soft approximation. In other words, if we take the resummed cross section in Mellin space, expand it in powers of $\as$ and then transform back in physical space term by term, we obtain the all-order cross section. This cross section has a divergent perturbative expansion like we saw in Section~\ref{sec:landaupole}.

Since the resummed cross section has the form of an exponentiation, it is easier to work in terms of the physical anomalous dimension, defined in Eq.~\eqref{eq:anomalousdim}. The physical anomalous dimension has the same divergent behaviour than the cross section. In particular, the order at which the series starts to diverge is independent of the logarithmic accuracy,
 so that we can study the behaviour of $\gamma_{\scriptstyle \rm LL}$, which is much simpler than the higher logarithmic orders. We recall that the series we want to study is Eq.~\eqref{eq:PasymptLL}
\be
R(\as(Q^2),x)=
\sum_{n=0}^\infty\Delta^{(n)}(1)\,[-\beta_0 \as(Q^2(1-x))]^{n+1},
\label{eq:landaupoleonly}
\ee
where $Q$ can be taken as the hard scale of the process (the Higgs or twice the top mass), and $x$ is what we usually call $\tau$:
\begin{equation}
\tau = \frac{M_H}{s},
\label{eq:taudef}
\end{equation}
the ratio between the invariant mass of the final state and the hadronic center-of-mass energy.

Once that we have determined the trend of the perturbative series, our theoretical uncertainty is simply the difference between the resummed result and the truncated series. This time it is more advisable to use the most accurate result (N$^3$LL), since while the divergent behaviour is independent of the logarithmic accuracy, the value of the resummed cross section strongly depends on it, so we cannot make any semplifications.

Finally,~\cite{higgsn3loapprox} and~\cite{topn3loapprox} find that a better approximation of the fixed order result is obtained if we introduce certain subdominant logarithmic contributions. For the sake of clarity, we will not include those contributions in the soft-approximate series. We will use what is called $N$-soft in~\cite{higgsn3loapprox}, that is to say the simplest approximation based on $\mathcal{D}^{\log}(N)$ of Eq.~\eqref{eq:DkMP_def}.
\section{Renormalons only}
Now let us suppose that the dominant source of divergence is the renormalon divergence. There are two models that we can build based on renormalons. In the first one, which we shall call \emph{Naive Renormalon Model} (NRM), the cross section exhibits the typical factorial divergence as it is. 

The second model, the \emph{Exponent Renormalon Model} (ERM), is built starting from the assumption that the renormalon divergence can occur inside the Sudakov exponent. The physical interpretation behind this model is that we are considering a class of corrections not to the whole process, but only to the soft gluons that are emitted (e.g. fermion bubble chain along an emitted gluon line).

So far, our discussion of renormalons has been made at the parton level: the divergent series we consider is the partonic cross section in $N$ space. In Section~\ref{sec:partonicvshadronic} we will show how we can extend the model to hadron level results (much easier to work with) without any trouble. The tacit assumption is that the PDFs, which are a perturbative expansion as well, don't present any divergent feature, or their divergence is subdominant with respect to that of the partonic cross section.

From now on, we will suppose that the first perturbative orders of an hadronic inclusive cross section are known, so that the truncated N$^k$LO cross section is given by
\begin{equation}
\sigma_\text{N$^k$LO} = \sigma_0 \left(1 + \sum_{n=0}^{k-1} c_n \as^{n+1} \right).
\label{eq:knownxs}
\end{equation}
%
\subsection{Naive Renormalon Model}
Recall that when renormalons are dominant, then the Borel transform of the perturbative series has the form
\begin{equation}
B[R] (u) = \sum_{m=-\infty}^{\infty} \frac{K_m}{1-\frac um},
\end{equation}
where $u = \beta_0 t$.
Renormalons are poles located at $u=m$ for every integer $m$, but the values of the residues $K_m$ are unknown. There are some ways to determine  these residues from first principles, using the bubble diagram chains, but those methods rely on strong approximations. The exact determination of the residues would require an all-order calculation, which is precisely what we are trying to avoid.
Instead, our approach consists in exploiting what we know about the series (the first known coefficients), to extrapolate the first and most important residues by comparison. 

If we know the cross section Eq.~\eqref{eq:knownxs} up to N$^3$LO, and therefore we possess three pieces of information beyond LO, we can assume that the Borel transform of the series only has three poles:
\begin{equation}
B[R](u) = \frac{K_{m_1}}{1-u/m_1} + \frac{K_{m_2}}{1-u/m_2} + \frac{K_{m_3}}{1 - u/m_3},
\end{equation}
where $m_n$ are the integers representing the position of the renormalon poles in the $u$ axis. If this is the Borel transform, then the series has the form 
\begin{equation}
\sigma^\text{NRM} = \sigma_0 \left(1 + \sum_{n=0}^{\oo} r_n \as^{n+1} \right),
\end{equation}
with the coefficients $r_n$ given by
\begin{equation}
r_n = K_{m_1} \left(\frac{\beta_0}{m_1} \right)^n \,  n! + K_{m_2} \left(\frac{\beta_0}{m_2} \right)^n \,  n! + K_{m_3} \left(\frac{\beta_0}{m_3} \right)^n \,  n!.
\label{eq:modelforrn}
\end{equation}
Now we ask that the NRM predict the first known orders exacly and solve the linear system for $n \in [0,2]$ to determine the residues:
\bea
\begin{cases}
c_0 = K_{m_1} + K_{m_2} + K_{m_3}  \\
c_1 = K_{m_1} \displaystyle \frac{\beta_0}{m_1}+ K_{m_2} \frac{\beta_0}{m_2}+ K_{m_3} \frac{\beta_0}{m_1}  \\
c_2 = 2 K_{m_1} \displaystyle \left( \frac{\beta_0}{m_1}\right)^2+ 2 K_{m_2} \left(\frac{\beta_0}{m_2}\right)^2+ 2 K_{m_3} \left(\frac{\beta_0}{m_1}\right)^2,
\label{eq:systemforK}
\end{cases}
\eea
where $c_i$ are the coefficients of Eq.~\eqref{eq:knownxs}.
Once that we have the residues, we know the perturbative series at all orders in $\as$ and we can easily see where the series begins to diverge. 

In order to evaluate its asymptotic value, recall that in Section~\ref{sec:Borel} we defined the Borel integral as~\eqref{eq:borelintegral}, so that in our case we can assign the following value to the sum of the series:
\bea
\tilde \sigma^\text{NRM} &=& \sigma_0 \left( 1+ \sum_{i=1}^3 K_{m_i}\, m_i \int_0^\oo \d t \, \frac{e^{-\frac{t}{\as}}}{m_i - \beta_0 t} \right)= \nonumber \\
&=& \sigma_0 \left[ 1 + \sum_{i=0}^3 \frac{K_{m_i} m_i}{\beta_0} \exp \left(-\frac{m_i}{ \beta_0 \as}\right) \, \text{Ei}\left(\frac{m_i}{ \beta_0 \as}\right) \right].
\label{eq:asymptoticNRM}
\eea
%
\subsection{Exponent Renormalon Model}
The ERM works exactly as the NRM, with the only difference that the renormalon series exponentiated. Let us take as an example the Higgs cross section. Recall that the resummed cross section has the form:
\begin{equation}
\hat \sigma^{\textrm{res}}( \as) = \hat \sigma_0 \, g_0(\alpha_s) \exp E(\as).
\label{eq:ERNmodel}
\end{equation}
In the ERM, the exponent has the typical renormalon form
\begin{equation}
E(\as) = \sum_{n=0}^\oo r_n \as^{n+1},
\label{eq:expERM}
\end{equation}
with $r_n$ given by Eq.~\eqref{eq:modelforrn}. To extract the values of the residues, we expand Eq.\eqref{eq:ERNmodel} in powers of $\as$:
\bea
\hat \sigma^{\textrm{res}}( \as) &=& \hat \sigma_0 \Bigg\{ 1 + (g_{01} + K_{m_1} + K_{m_2} + K_{m_3})\, \as + \nonumber  \\
&+& \left[g_{02} + g_{01} (K_{m_1} + K_{m_2} + K_{m_3}) + 
 \frac 12 ((K_{m_1} + K_{m_2} + K_{m_3})^2 + \right. \nonumber \\
    &+& \left. 2 \left(\frac{K_{m_1}}{m_1} \beta_0 + \frac{K_{m_2}}{m_2} \beta_0 + \frac{K_{m_3}}{m_2} \beta_0\right)\right]\as^2 + \dots \Bigg\},
\eea
and then we confront order by order with the exact cross section to extrapolate the residues $K_{m_1}$, $K_{m_2}$ and $K_{m_3}$.

Once again, using the residues we are able to know the series at all orders. The asymptotic sum of the series is nothing but the exponential of the Borel integral:
\begin{equation}
\tilde \sigma^\text{ERM} = \sigma_0 \, g_0(\as) \, \exp\left[ \sum_{i=0}^3 \frac{K_{m_i} m_i}{\beta_0} e^{-\frac{m_i}{ \beta_0 \as}} \, \text{Ei}\left(\frac{m_i}{ \beta_0 \as}\right) \right].
\end{equation}
%
\subsection{Partonic vs hadronic cross section}
\label{sec:partonicvshadronic}
We have formulated our models at the hadronic level, but to be more precise we should have applied our models to the partonic cross section in $N$ (or $z$) space. In fact, for every $N$ there should be a renormalon series with residues $K_i (N)$. We will now show that we can apply both renormalon models directly to the hadronic cross section, without going through $N$ space. The basic idea is that once we perform the inverse Mellin integral to obtain the hadronic cross section, the $N$-dependence is converted to a $\tau$-dependence. But $\tau$~\eqref{eq:taudef} is fixed by the invariant mass of the final state and by the center-of-mass energy of the collision. Hence, the residues are pure numbers.

Let us consider the NRM with only one renormalon for simplicity. The generalization to the case of multiple renormalons and to the ERM is straightforward. Remember that the inclusive hadronic cross section, differential only in $M$, can be written as the inverse Mellin transform of the cross section in $N$ space:
\begin{equation}
\sigma(\tau,M^2) =  \frac{1}{2 \pi i} \int_{c-i\infty}^{c+i\infty} dN \ \tau^{-N} \sigma(N),
\end{equation}
where the hadronic cross section in $N$ space  $\sigma(N)$ is given by
\begin{equation}
\sigma(N) = \hat \sigma_0 (\as) \, \Lum(N) \left[K(N) \sum_{n=0}^\oo \left(\frac{\beta_0}{m}\right)^n n! \right].
\end{equation}
Now, once we perform the Mellin inversion integral we obtain
\begin{equation}
\sigma(\tau,M^2) = \sigma_0 (\as) \sum_{n=0}^\oo  \left(\frac{\beta_0}{m}\right)^n n!  \left[\frac{1}{2 \pi i} \int_{c-i\infty}^{c+i\infty} dN \ \tau^{-N}  \Lum(N) \, K(N)\right].
\end{equation}
Therefore, the residue at hadronic level can be defined as 
\begin{equation}
K \equiv \frac{1}{2 \pi i} \int_{c-i\infty}^{c+i\infty} dN \ \tau^{-N}  \Lum(N) \, K(N).
\end{equation}
The same exact reasoning holds for the physical space (although the coefficient functions are distributions) and, with small modifications, for the ERM.
\section{Renormalons and Landau pole divergence}
Since the soft approximation is particularly good at all the known orders, we can make the assumption that the series is dominated by soft terms at very low perturbative orders, and the renormalon effects start to kick in later. In this scenario, the renormalons would cover the discrepancy between the soft approximation and the exact result at low orders. Once again, we divide the case in which the renormalons are outside or inside the Sudakov exponent.
\subsection{Naive Soft-Renormalon Model}
Let us consider a situation where the hadronic cross section is simply the sum of a soft and a renormalon contribution:
\begin{equation}
\sigma = \sigma^{\text{soft}} + \sigma^{\text{ren}},
\end{equation}
where
\begin{equation}
\sigma^{\text{soft}} = \sigma_0 \left(1 + \sum_{n=0}^\oo c^\text{soft}_n \as^{n+1} \right)
\label{eq:sigmasoft}
\end{equation}
is the soft approximation to the exact cross section based on $\mathcal{D}^{\log}(N)$ of Eq.~\eqref{eq:DkMP_def}.
More specifically, considering the N$^k$LO cross section of Eq.~\eqref{eq:knownxs}, we can separate the soft and the renormalon part:
\begin{equation}
\sigma_\text{N$^{k}$LO} =
\sigma_0 \left[1 +  \sum_{n=0}^{k-1}  c_n^\text{soft} \alpha_s^{n+1}+ \sum_{n=0}^{k-1} c_n^\text{ren}  \alpha_s^{n+1} \right]=
\sigma_\text{N$^{k}$LO}^\text{soft} + \sigma_0 \sum_{n=0}^{k-1} c_n^\text{ren} \alpha_s^{n+1}.
\label{eq:ESRM}
\end{equation}
We now compare the renormalon coefficients $r_n$ of Eq.~\eqref{eq:modelforrn} with the renormalon coefficients $c_n^\text{ren}$ of Eq.~\eqref{eq:ESRM}, that we can determine recursively from $\sigma$ and $\sigma^\text{soft}$ using
\begin{equation}
c_k^\text{ren} = \frac{1}{\sigma_0 \, \alpha_s^{k+1}} \left(\sigma_\text{N$^{k+1}$LO} - \sigma^\text{soft}_\text{N$^{k+1}$LO} - \sigma_0 \sum_{n=0}^{k-1} c^\text{ren}_n \alpha_s^{n+1} \right).
\end{equation}

For what concerns the asymptotic value, the sum of the soft part of the series is the resummed NNLL result computed with the Borel prescription described in Section~\ref{sec:borelprescription}. Note that since we are interested in the soft contribution only we shall not use the NNLL+NNLO result like in Eq.~\eqref{eq:matching}. We denote as $\sigma^\text{res}_\text{NNLL}$ the resummed $N$-soft cross section~\eqref{eq:DkMP_def} without fixed-order matching.
The sum of the renormalon part, instead, is the usual Borel integral. Therefore, the sum of the whole series in the Naive Soft Renormalon Model is
\begin{equation}
\tilde \sigma^\text{NSRM} = \sigma^\text{res}_\text{NNLL} + \sigma_0 \left[ \sum_{i=0}^3 \frac{K_{m_i} m_i}{\beta_0}e^{-\frac{m_i}{ \beta_0 \as}}\, \text{Ei}\left(\frac{m_i}{ \beta_0 \as}\right) \right].
\end{equation}
%
\subsection{Exponent Soft-Renormalon Model}
The equivalent of the ERM that fully includes soft contributions is a model where the cross section has the form
\bea
\sigma^\text{ESRM} &=& \sigma_0 \, g_0 (\as) \exp \left[ E^\text{soft}(\as) + E^\text{ren}(\as) \right] = \sigma^\text{soft} \, \exp \left[  E^\text{ren}(\as) \right] = \nonumber \\
&=& \sigma_0 \left(1 + \sum_{n=0}^\oo c^\text{soft}_n \as^{n+1} \right) \exp \left[ \sum_{n=0}^\oo r_n \as^{n+1}\right].
\label{eq:ESRMxs}
\eea
where we used the renormalon exponent $E^\text{ren}$ of Eq.~\eqref{eq:expERM} and $\sigma^\text{soft}$  of Eq.~\eqref{eq:sigmasoft}. To compute the residues, we need to compare the fixed order results with the perturbative expansion of Eq.~\eqref{eq:ESRMxs}, which is
\bea
\sigma^\text{ESRM} &=& \sigma_0 \Bigg\{ 1+ \left[K_{m_1} + K_{m_2} + K_{m_3} + c^\text{soft}_1\right] \as + \nonumber \\
&+& \left[\frac 12 \left(K_{m_1} + K_{m_2} + K_{m_3} \right)^2 + 
       \left(K_{m_1} \frac{\beta_0}{m_1} + K_{m_2} \frac{\beta_0}{m_1} +K_{m_3} \frac{\beta_0}{m_1}\right) +\right. \nonumber \\
       &+& \left. (k1 + k2 + k3) c_1^\text{soft} + c_2^\text{soft}\right] \as^2 + \dots \Bigg\}. 
\eea
Then, the asymptotic sum of the series is
\begin{equation}
\tilde \sigma^\text{ERM} = \sigma^\text{res}_\text{NNLL} \, \exp\left[ \sum_{i=0}^3 \frac{K_{m_i} m_i}{\beta_0} e^{-\frac{m_i}{ \beta_0 \as}} \, \text{Ei}\left(\frac{m_i}{ \beta_0 \as}\right) \right].
\end{equation}
%
\section{Divergence point $\bar n$}
\label{sec:divergencepoint}
In this Section we analyze the trend of the  partial sum in each of the models, in order to see at which perturbative order the divergence occurs. We will see that the Landau pole divergence 
occurs much later than the renormalon divergence. We will also show that the divergence point in any of the renormalon models only depends on the location of the leading renormalon pole. It goes without saying that in the models with both sources of divergence, the renormalon divergence prevales.
\subsection{Landau pole divergence only}
Let us recall the form of $\gamma_{\scriptsize \rm LL}$ Eq.~\eqref{eq:landaupoleonly} and write its partial sum $R_k$:
\begin{equation}
R_k=\sum_{n=0}^k s_n =\sum_{n=0}^k \Delta^{(n)}(1)\,[-\beta_0 \as(Q^2(1-x))]^{n+1}, \nonumber
\end{equation}
 where $\Delta(z) \equiv 1/\Gamma(z)$. This series represents the physical anomalous dimension at LL. The physical anomalous dimension is related to the exponent of the resummed cross section, and it is obvious that if a series has the form $\sigma_n \propto \exp(E_n)$, then $\sigma_n$ starts to diverge when $E_n$ starts to diverge.
\begin{figure}
\centering
\includegraphics[width=0.8\textwidth]{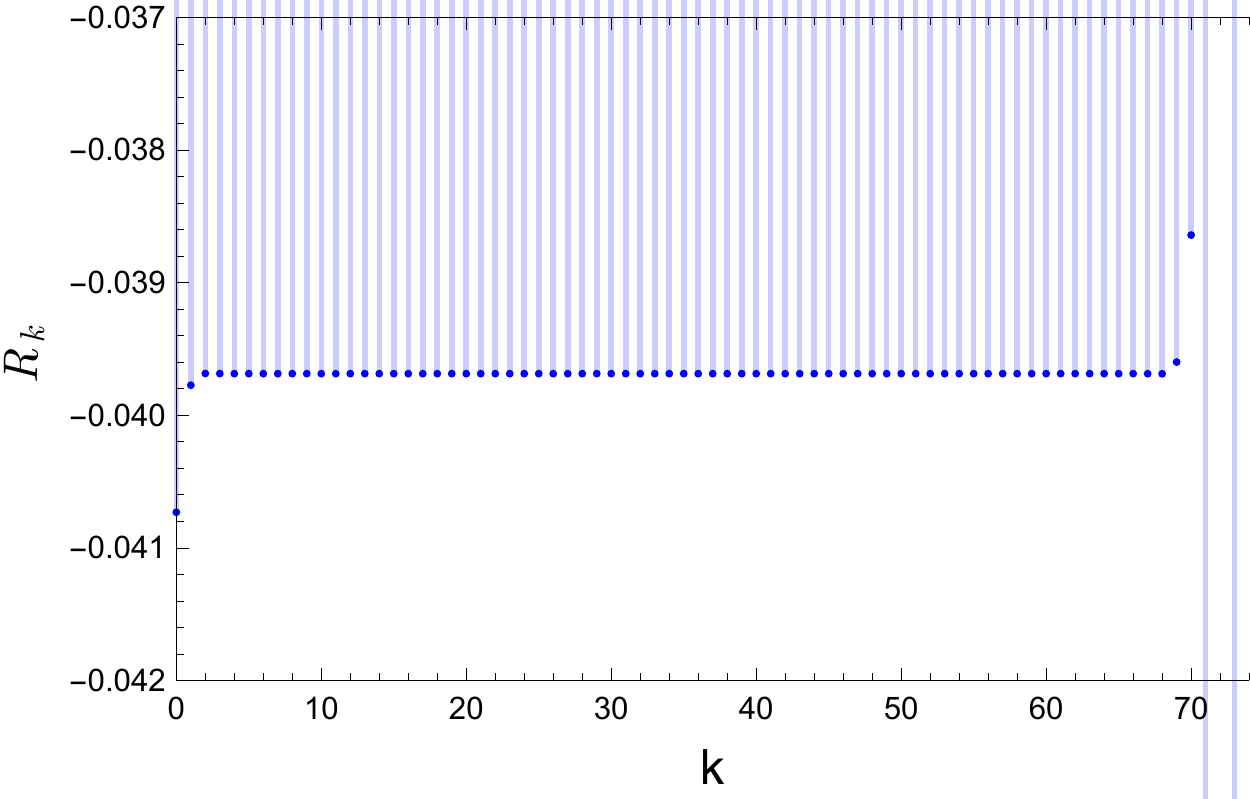} \\
\centering
\includegraphics[width=0.8\textwidth]{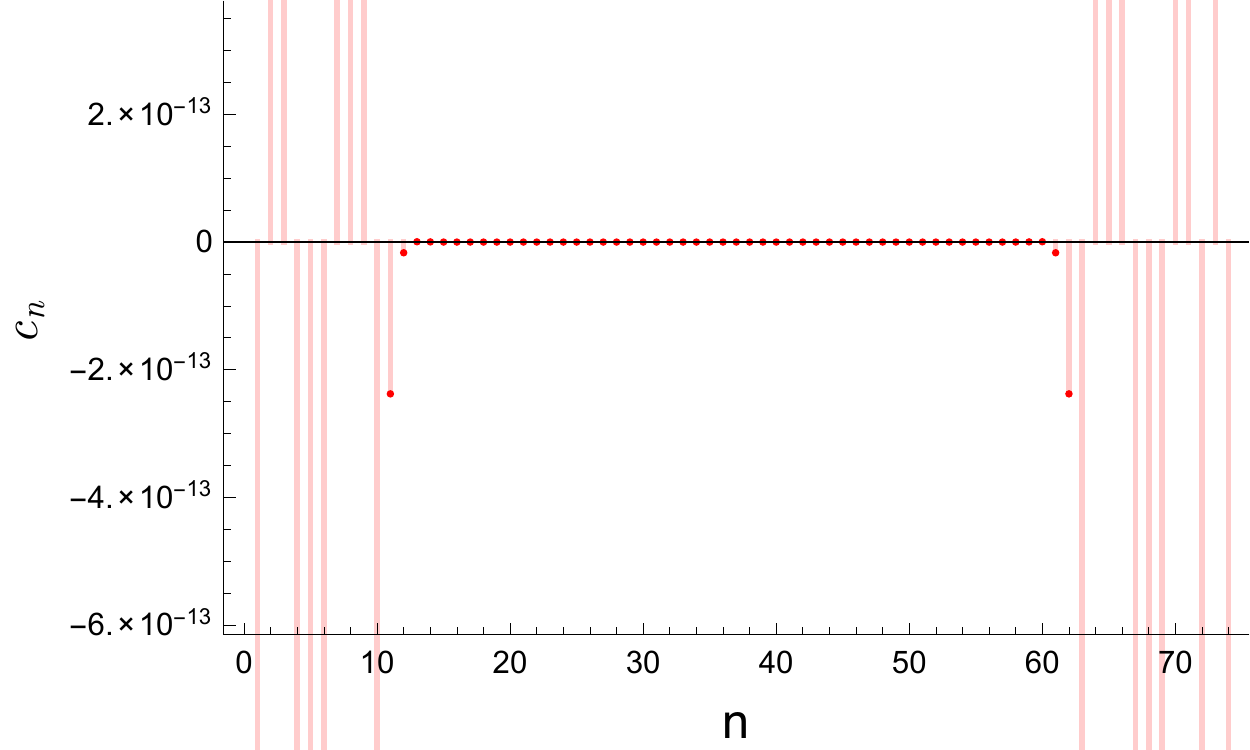}
\caption{Trend of the partial sum $R_k$ (top, blue) and of the sequence $s_n$ (bottom, red), supposing that only the Landau pole divergence exists. Results are obtained using $x = \frac{m_H^2}{s}$, with $s=13\text{ TeV}$, and are almost identical if we use the $t \bar t$ invariant mass instead of the Higgs mass.}
\label{fig:landaupoleonly}
\end{figure}

In Fig.~\ref{fig:landaupoleonly} the trend of the partial sum $R_k$ and of the sequence of contributions $s_n$ is shown. We see that the series converges rapidly to a plateau (the quantity that is summed at each perturbative order is very small), and then the partial contributions start to gain a noticeable value at very high perturbative orders ($k \sim 70$). One usually defines the perturbative order $\bar n$ at which the series starts to diverge by imposing that
\begin{equation}
|s_n| > |s_{n-1}| \qquad \text{for} \quad n> \bar{n}.
\end{equation}
With this definition, we find that $\bar n = 33$. Note that this does not imply that the series is already divergent at $\bar n$. It only means that from that point on the contributions that we are adding start to grow. However, the macroscopic effect is only visible at $n\sim 70$.

The result only depends on $x$, which we take as $\tau$ of Eq.~\eqref{eq:taudef}. The $x$ value corresponding to the worst behaviour of $\gamma_{\scriptsize \rm LL}$ is $x\sim1$. However, for the Higgs production at 13 TeV center-of-mass energy $x \sim 10^{-4}$. Therefore, we conclude that there is no reason to worry about the Landau pole divergence since, for processes far away from threshold, the divergence kicks in at very high perturbative orders.
\subsection{Renormalon divergence only}
\label{sec:renormnbar}
For both models that involve the renormalon divergence, the order at which the series starts to diverge $\bar n$ depends on the location of the leading renormalon. Before plotting the outcome of the models, we will justify this statement analytically. In our models multiple poles are present but, as we are going to show,  the dominant pole is the one that is closer to 0. Therefore, let us suppose that the renormalon series has only one pole in its Borel transform and hence only one residue. The partial sum in this scenario is
\begin{equation}
R_n = K \sum_{k=0}^n (a \beta_0 )^k\, k! \,\as^{k+1},
\end{equation}
where $a=1/m$ is the inverse of the location of the pole in the $u$ axis. To prove our point, we shall compute the increment and study its trend with respect to $n$:
\begin{equation}
R_{n+1}-R_n = K a \beta_0 \as^2 \, (a \beta_0 \as)^n\, \Gamma(n+2).
\end{equation}

If we continue this function analytically for real $n$ and find a minimum at $n_0$, it means that the contributions that we are adding to the partial sum are decreasing before $n_0$, and start increasing afterwards. To look for a minimum, we compute the derivative of the increment with respect to $n$ and put it to 0.
\begin{equation}
\frac{\d}{\d n} \left[ (a \beta_0 \as)^n\, \Gamma(n+2) \right]= 0 \implies \log(a \beta_0 \as) + \psi(n+2) =0,
\label{eq:divergencepoint}
\end{equation}
where $\psi(z)$ is the digamma function, the logarithmic derivative of $\Gamma(z)$. We could solve this equation numerically, but a very interesting formula comes out if we use the Stirling approximation, supposing that $n$ is big, as can be checked a posteriori. In fact, if we define $t=n+1$,
\bea
\psi(t+1) = \frac{\d}{\d t} \log \Gamma(t+1) \sim \frac{\d}{\d t} \log \left[ \left(\frac te \right)^t \sqrt{2 \pi t} \right] = \frac{\d}{\d t} \left[t \log t - t + \frac 12 \log t \right] = \log t + \frac{1}{2 t}. \nonumber \\
\eea
Now, since we are considering large $t$
\begin{equation}
\frac{1}{2 t} \sim \log \left( 1 + \frac{1}{2 t}\right),
\end{equation}
and therefore we have
\begin{equation}
\psi (t+1) \sim \log \left( 1 + \frac{1}{2 t} \right).
\end{equation}
The equation which determines the minimum becomes therefore
\begin{equation}
\log \left( 1 + \frac{1}{2 t} \right) = \log \left( \frac{1}{a \beta_0 \as} \right)  \implies \bar n = \frac{1}{a \beta_0 \as} - \frac{3}{2},
\end{equation}
where $m$ denotes the location of the pole in the Borel transform. Numerical solutions of Eq.~\eqref{eq:divergencepoint} for integer values of $1/a$ are in good agreement with this approximation.

We have thus proven that the divergence point is directly proportional to the position of the leading renormalon. There are arguments~\cite{beneke} in favour of the hypothesis that, for processes like Drell-Yan, Higgs and $t \bar t$ production, the leading renormalon is located at $m=1$. These arguments are based on the fact that a previously supposed pole at $m=1/2$ is in reality canceled at all orders. Hence, we conclude that the renormalon divergence occurs much before than the Landau pole divergence, more precisely at
\begin{equation}
\bar n \sim 13.
\end{equation}
Finally, we note that Eq.~\eqref{eq:divergencepoint} has the same solution if we consider the ERM instead of the NRM, as one could easily expect.

One last observation can be made about the instanton divergence, justifying what has been said in Section~\ref{sec:instanton}. Since the location of the leading instanton pole in the Borel plane is at $t = 4 \pi$, the instanton divergence occurs at $\bar n \sim 110 $, therefore much later than the Landau pole divergence.
\section{Theoretical uncertainties}
This Section contains our predictions for the theoretical uncertainties on the perturbative series in QCD,  based on the models that we introduced above. In each of the models, the difference between the all-order result and the last known order can be used as an estimate for the theoretical uncertainty. However, there is a certain arbitrariness in determining the asymptotic sum of the series, i.e. the freedom in choosing the renormalon poles whose residues are extrapolated by comparison from the exact known orders.

We will consider in particular the process of Higgs production in gluon fusion and, in Section~\ref{sec:top}, $t \bar t$ production in gluon fusion.
Throughout this Section, for phenomenology we will use the code \texttt{ihixs 1.4}~\cite{ihixs} to obtain the inclusive Higgs cross section up to NNLO at a scale $\mu_R = \mu_F= m_H$. We  will then add the exact N$^3$LO result that can be found on~\cite{anastasiou}. This last result is obtained using the PDF set MSTW2008nnlo68cl, so we will consistently use that same PDF set as an input for the code. Since we are trying to neglect all the PDF perturbative effects, we will always use the most accurate set available, i.e. NNLO. The impact of theoretical uncertainties on PDFs on the Higgs cross section is negligible~\cite{doweneed}.

We are going to need the soft-gluon approximation to the Higgs cross section $N$-soft, described in Section~\ref{sec:softgluonapprox}. To compute it, we use  \texttt{ggHiggs 2.1}\cite{gghiggs}, while to compute the resummed result we use  \texttt{TROLL}~\cite{troll} with the $N$-soft prescription. Remember that the soft approximation described in~\cite{higgsn3loapprox} also contains some sub-leading contributions, but for the sake of clarity we are not going to take them into account. We will use the simpler approximation based on $\mathcal{D}^{\log}(N)$ as in Eq.~\eqref{eq:DkMP_def}.

For $t \bar t$ production, we will use \texttt{Top++ 2.0}~\cite{top++} with NNPDF3.0 PDFs. The $N$-soft approximation, discussed in Ref.~\cite{topn3loapprox}, will be extracted from a private code by Claudio Muselli.
\subsection{Choice of the poles}
Like we said before, for the Higgs case only three orders are known beyond the LO, so we have a system with 3 equations and 3 unknowns residues. The location of the poles of Eq.~\eqref{eq:modelforrn}, however,  is free. For example, an obvious choice would be $m_1 = 1$, $m_2 = 2$ and $m_3 = 3$, but then our model would not predict the behaviour induced by the UV renormalons, which are located in the negative $u$ axis.

If we compute the asymptotic sum of the series for different choices of the poles, we notice that the maximum value $\Sigma_\text{max}$ is obtained for $\begin{pmatrix}m_1 & m_2 & m_3 \end{pmatrix}= \begin{pmatrix}1 & 2 & 3 \end{pmatrix}$, and the minimum $\Sigma_\text{min}$ for $\begin{pmatrix}-1 & -2 & -3 \end{pmatrix}$, all the other values being included inside that interval. This is not easy to prove, but one can solve the system~\eqref{eq:systemforK} for generic $m_i$ and see that the results of the Borel integral Eq.~\eqref{eq:asymptoticNRM} obey to this rule.
\begin{table}
\caption{Values of the Borel integral in the NRM for the Higgs cross section at 13 TeV, pdf set MSTW2008nnlo68cl. Results up to NNLO obtained with \texttt{ihixs}, the N$^3$LO is taken from~\cite{anastasiou}.}
\centering
\begin{tabular}{cc}
\toprule
Pole choice $\begin{pmatrix}m_1 & m_2 & m_3\end{pmatrix}$ & Borel integral [pb] \\
\midrule
$\begin{pmatrix}1 & 2 & 3 \end{pmatrix}$ & 44.5645 \\
$\begin{pmatrix}1 & 3 & 4 \end{pmatrix}$ & 44.419 \\
$\begin{pmatrix}2 & 3 & 4 \end{pmatrix}$ & 43.8916 \\
$\begin{pmatrix}-2 & -3 & -4 \end{pmatrix}$ & 42.4534 \\
$\begin{pmatrix} -1& -3 & -4 \end{pmatrix}$ & 42.1961 \\
$\begin{pmatrix}-1 & -2 & -3 \end{pmatrix}$ & 42.0315 \\
\bottomrule
\end{tabular}
\label{tab:polechoice}
\end{table}

In Table~\ref{tab:polechoice}, as an example, the asymptotic sum of the Higgs cross section is computed using different pole choices in the NRM.
\subsection{Trend of the perturbative expansion}
\begin{figure}
\centering
\includegraphics[width=0.9\textwidth]{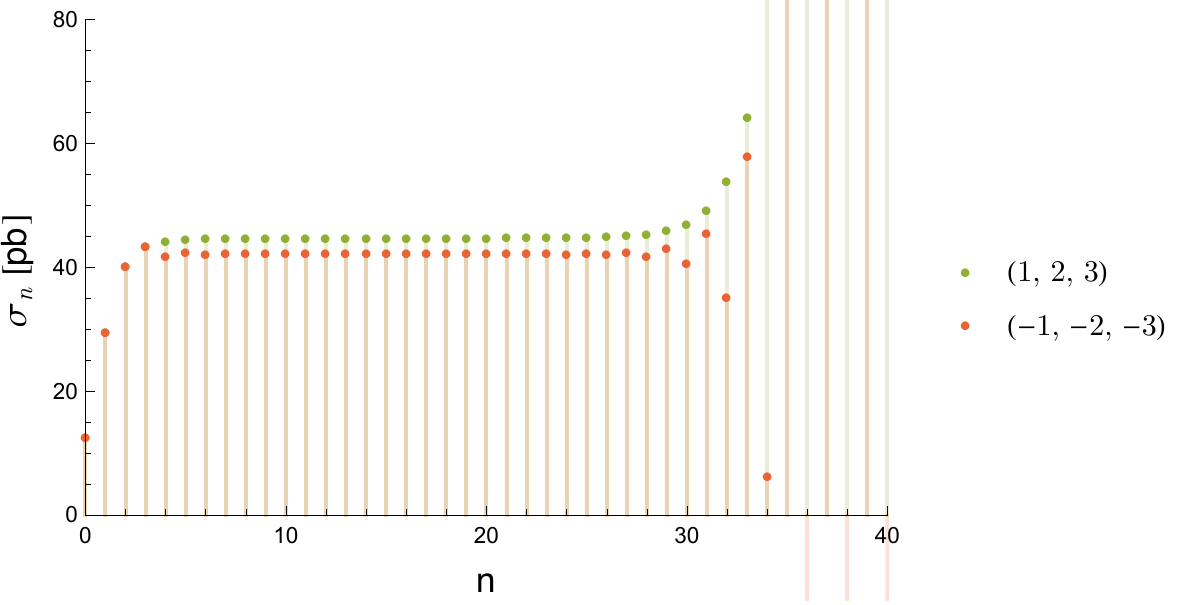}
\caption{Trend of the partial sum as function of the perturbative order for the Higgs cross section at 13 TeV center-of-mass energy in the NRM. Results up to NNLO are obtained with \texttt{ihixs}, the N$^3$LO is taken from~\cite{anastasiou},  PDF set MSTW2008nnlo68cl.}
\label{fig:NRMtrend}
\end{figure}
In Fig.~\ref{fig:NRMtrend} the partial sum of the series $\sigma_n$ in the NRM is plotted as a function of the perturbative order $n$. The different colors represent different choices of the poles, in particular we plot the one that maximizes and the one that minimizes the asymptotic sum of the series. We note that the series exhibits a fixed-sign divergence when IR renormalons are present, while it exhibits an alternating-sign divergence when UV renormalons are present, like we studied in Chapter~\ref{ch:renormalons}.

We have chosen the NRM as en example: plots like Fig.~\ref{fig:NRMtrend} are similar in all the models. What really changes between different models is the estimate for the theoretical uncertainty on the last known order. In the next Section, we will compare the theoretical uncertainty estimates between different models.
\subsection{Comparison between our models}
Let us summarize how we can study the higher-order behaviour of a perturbative expansion in QCD. In the NRM, we assume that the series has a simple factorial divergence. In the ERM, the factorial behaviour is exponentiated. The NSRM and ESRM are the extensions of respectively the NRM and the ERM when we take into account the soft-gluon approximation.

The trend of the partial sum does not depend strongly on the model: the leading renormalon pole is always located at $|u|=1$ and therefore the divergence occurs always at $\bar n \sim 13$, like we saw in Section~\ref{sec:renormnbar}. In fact, even when the Landau pole divergence is present together with the renormalon divergence, the latter is dominant.

The freedom to choose the poles is also common to each of our models. In particular, it is always true that the maximum and minimum values for the asymptotic sum of the series correspond to the choice of the poles $(1,2,3)$ and $(-1,-2,-3)$, although maximum and minimum can be exchanged. What matters is that any other pole choice leads to an asymptotic sum included in that range.

Once that we have the truncated sum $\sigma_k$ (the N$^k$LO) and the asymptotic value $\Sigma$, we say that the difference between the two is our theoretical uncertainty. However, we do not have a single asymptotic value, but a range of values $[\Sigma_\text{min}, \Sigma_\text{max}]$. Our uncertainty interval $\Delta$, then, is defined as
\begin{equation}
\Delta = 
\begin{cases}
[\Sigma_\text{min}, \Sigma_\text{max}] \qquad & \text{if } \sigma_k \in [\Sigma_\text{min}, \Sigma_\text{max}]  \\
[\Sigma_\text{min}, \sigma_k] \qquad & \text{if } \sigma_k > \Sigma_\text{max} \\
[\sigma_k, \Sigma_\text{max}] \qquad & \text{if } \sigma_k < \Sigma_\text{min}
\end{cases}
\end{equation}

\begin{figure}
\centering
\includegraphics[width=0.8\textwidth]{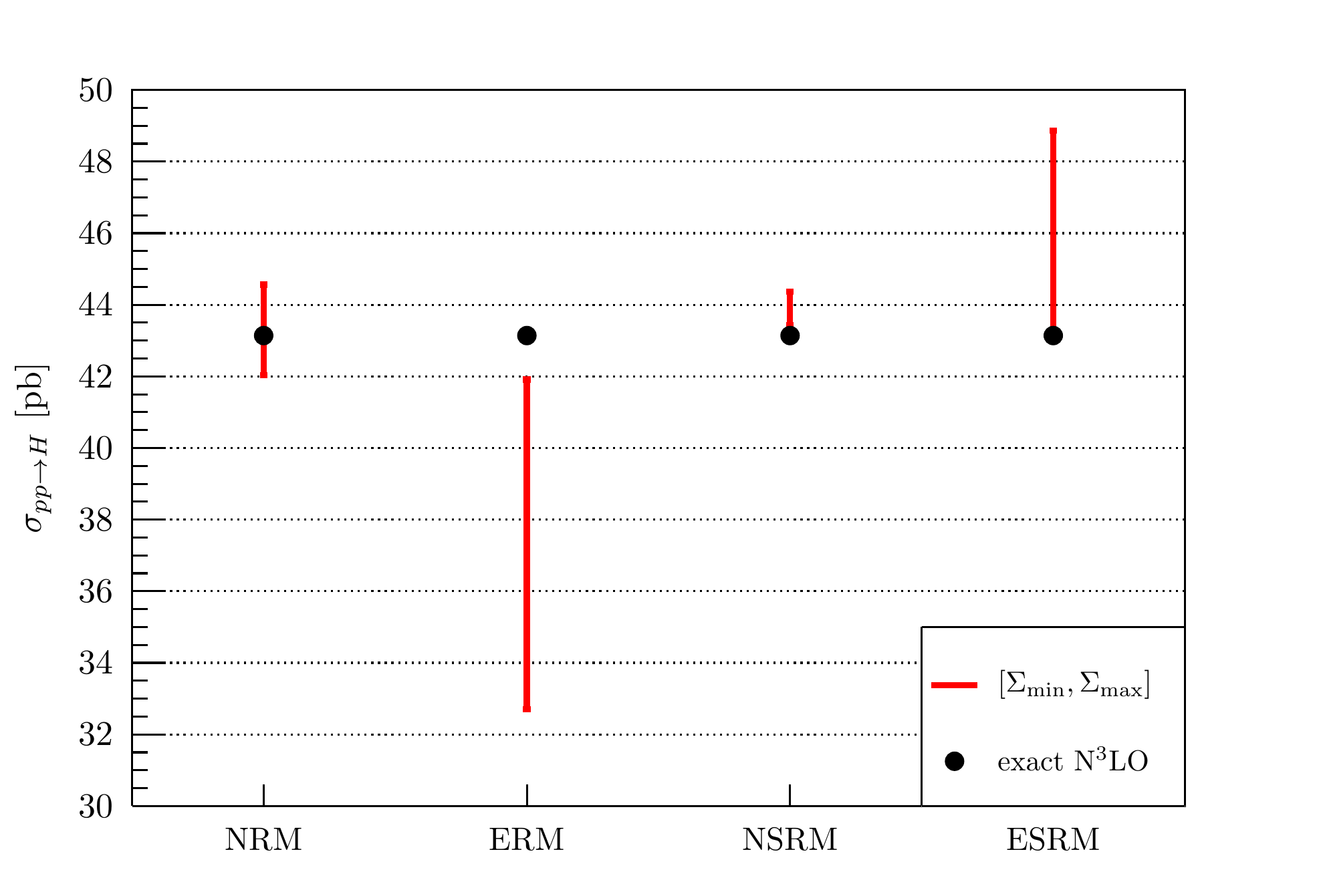}
\caption{ Higgs cross section: intervals for $[\Sigma_\text{max},\Sigma_\text{min}]$, the maximum and minimum value for the asymptotic sum of the series, together with the exact N$^3$LO. Results from different models are compared. We used \texttt{ihixs} and~\cite{anastasiou}, finite top mass and finite bottom mass effects are included and the pdf set is MSTW2008nnlo68cl. }
\label{fig:mymodelcomp}
\end{figure}

Now, what really makes our models different from one another is precisely the value of the predicted $\Sigma_\text{max}$ and $\Sigma_\text{min}$. In Fig.~\ref{fig:mymodelcomp}, these asymptotic values are computed in each of our models and compared. At a first glance, we notice that the Exponent Models tend to give a wider spread of possible values for the asymptotic sum $\Sigma$, and therefore will have a broader theoretical uncertainty band than the Naive Models.
%
\section{Accuracy of the models}
We have introduced 4 different models for the theoretical uncertainty on the Higgs cross section, but their results don't seem to be in good agreement. In order to determine which of them give accurate predictions we perform two test. First we see which models describe the uncertainty on the NNLO correctly, then we compare our estimates with the  estimates predicted by all the other models described in Chapter~\ref{ch:theorunc}.

\subsection{Accuracy in predicting the N$^3$LO}
\begin{figure}
\centering
\includegraphics[width=0.8\textwidth]{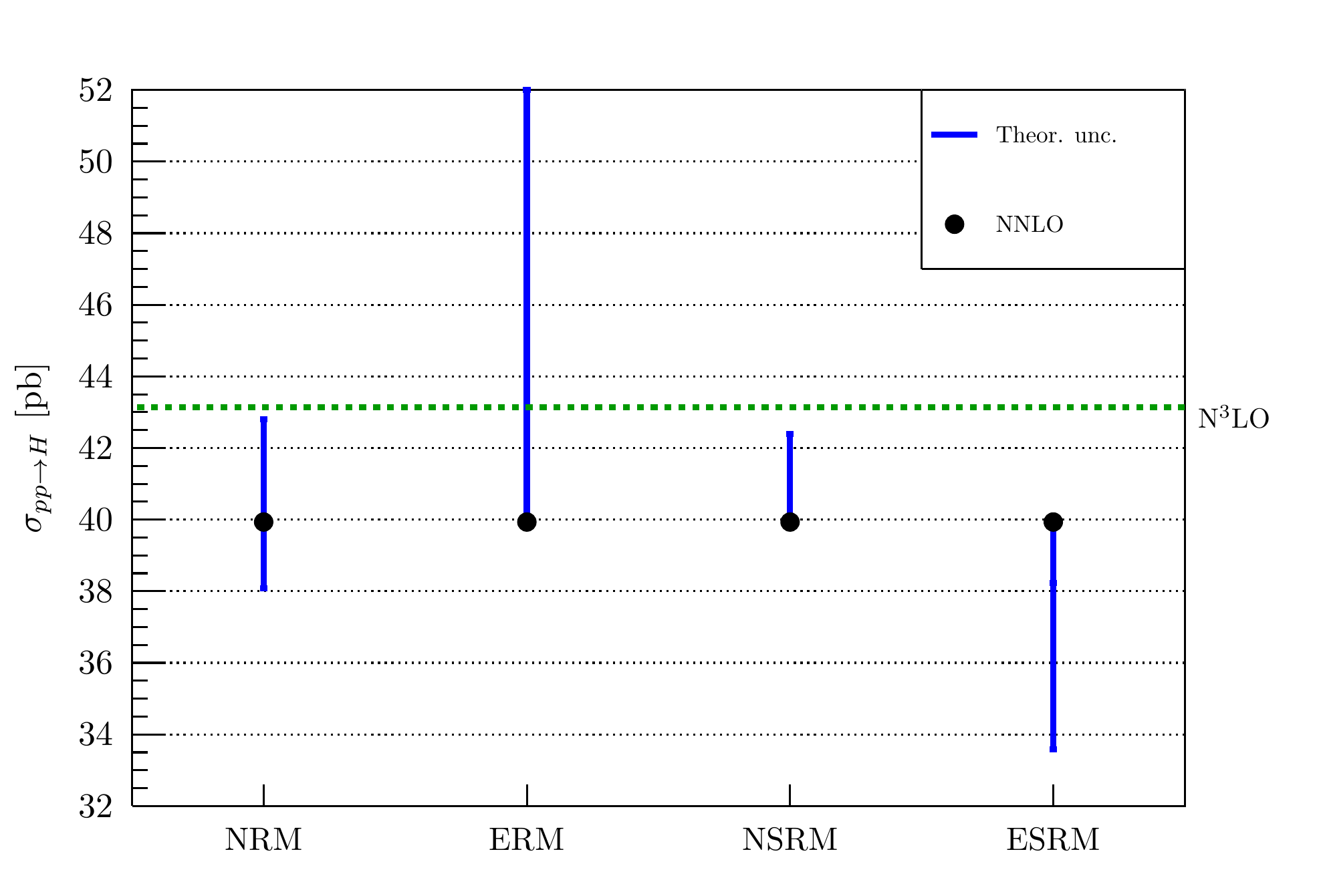}
\caption{Theoretical uncertainty bands on the NNLO Higgs cross section (blue lines), computed pretending not to know the N$^3$LO. The exact N$^3$LO is displayed as a dashed green line.}
\label{fig:myn3lohiggs}
\end{figure}
Let us pretend that we do not to know the N$^3$LO nor the resummed N$^3$LL cross section. We can apply our methods to the truncated partial sum up to NNLO and see if the uncertainty bands include the true N$^3$LO or not.

If we know the cross section up to NNLO, there are only two residues that we can fit, and therefore all the models suffer from lack of precision. The $\Sigma_\text{max}$ and $\Sigma_\text{min}$ values are obtained using $(1,2)$ and $(-1,-2)$ as poles. In Figure~\ref{fig:myn3lohiggs}, results are plotted for the theoretical uncertainties on the NNLO. The plot also displays the actual value of the N$^3$LO.

We notice that the only model whose uncertainty band contains the N$^3$LO cross section is the Exponent Renormalon Model. However, the uncertainty band in the ERM is very large. The NRM and NSRM are not far from the true value, while the ESRM seems to completely miss it.

Furthermore, if we compare Fig.~\ref{fig:myn3lohiggs} and Fig.~\ref{fig:mymodelcomp}, we notice that the ERM and the ESRM have completely opposite predictions. In fact, if we include the N$^3$LO the ERM predicts an asymptotic sum which is smaller than the N$^3$LO itself. Without N$^3$LO, instead, the uncertainty band is on the upper side of the NNLO. The opposite thing happens for the ESRM.  We have checked that, due to the exponential,  the outcomes of the Exponent Models are highly unstable with minimal variations of the initial values of the cross section. For example, if we use the cross section up to NNLO computed by \texttt{ggHiggs}, we obtain a very short uncertainty band and the approximate N$^3$LO lies perfectly inside that band\footnote{The NNLO computed with \texttt{ggHiggs} differs from the NNLO computed with \texttt{ihixs} by some finite top mass effects.}.

We conclude that the Naive Models represent our best models for THU. They are both quite accurate at NNLO and stable with small variations of the initial conditions. Furthermore, if we compare Fig.~\ref{fig:mymodelcomp}  with Fig.~\ref{fig:myn3lohiggs}, we note that the uncertainty band on the N$^3$LO is smaller than the one on the NNLO, as one would expect from a convergent method. In particular, the NSRM predicts a positive uncertainty band, which we expect from the monotonicity of the perturbative cross section.
%
%
\subsection{Consistency with other models}
After checking the accuracy of our models at NNLO, we have ruled out the Exponent models since their predictions don't seem to be reliable. We would like now to compare the estimates for the THU in the NRM and in the NSRM with the other known models described in Chapter~\ref{ch:theorunc}.

There, we saw that the conventional way to give a theoretical error band is to vary the renormalization and factorization scale $\mu_R$ and $\mu_F$ around a central value (usually $\mu_R=\mu_F=m_H$). This model is called Scale Variation. The Cacciari-Houdeau model~\cite{cacciarihoudeau}, which we shall use in its modified version for hadronic observable~\cite{cacciaribagnaschi}, is instead based on a Bayesian approach and gives a 68\% confidence level interval for the theoretical uncertainty on higher orders. The David-Passarino model~\cite{davidpassarino} makes use of sequence transformations to improve the convergence of the series\footnote{The uncertainty given in~\cite{davidpassarino} is valid for the Higgs cross section at 8 TeV, here we have simply rescaled the uncertainty band for the 13 TeV cross section.}. 

\begin{figure}
\centering
\includegraphics[width=0.9\textwidth]{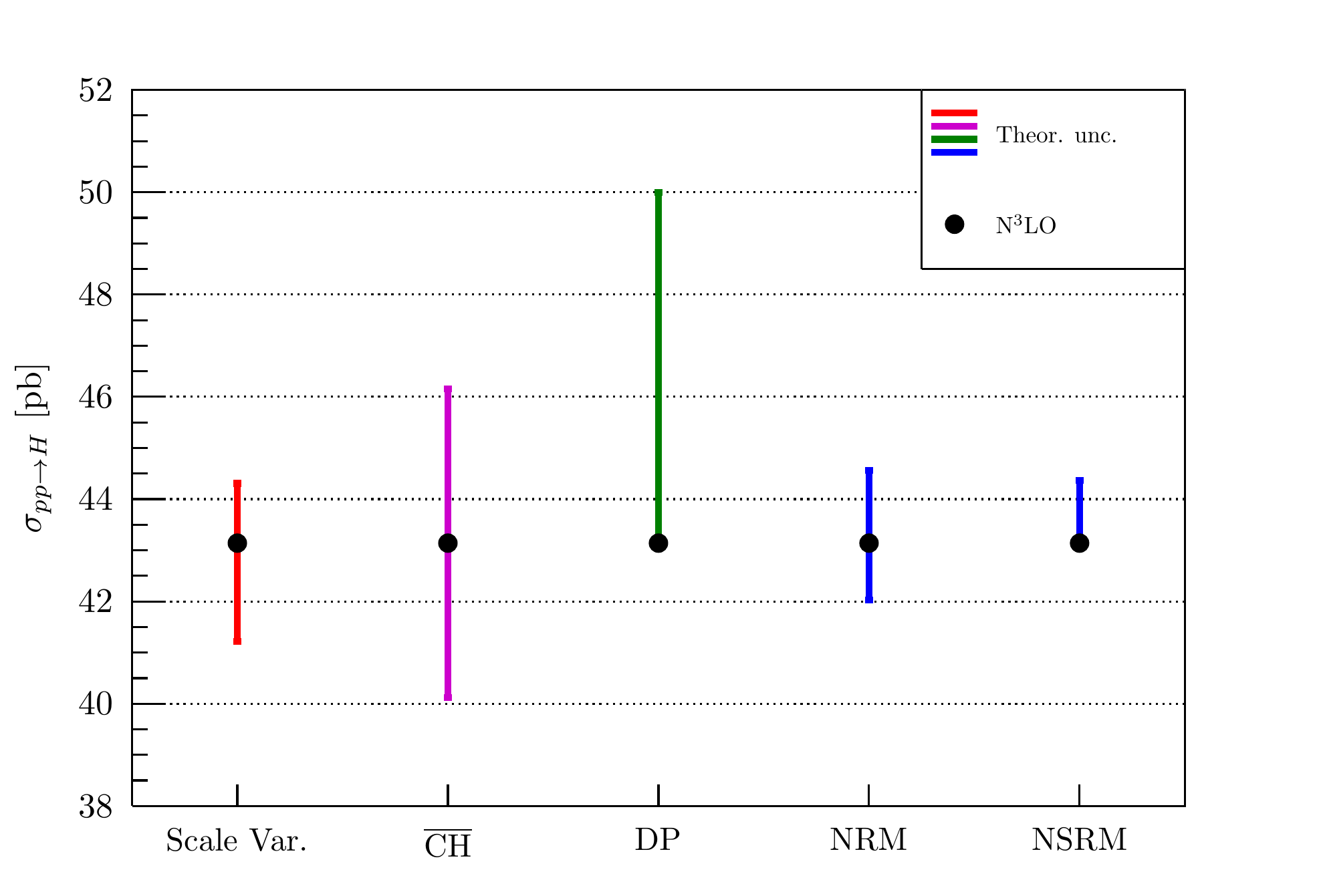}
\caption{Various models for the theoretical uncertainty on the N$^3$LO are considered. In red the Scale Variation uncertainty is displayed, in violet the Cacciari-Houdeau modified model $\overline{\text{CH}}$ with 68\% c.l., in green the David-Passarino uncertainty and in blue our models: the Naive Renormalon Model and the Naive Soft Renormalon Model.}
\label{fig:othermethodcomp}
\end{figure}

Fig.~\ref{fig:othermethodcomp} shows the comparison between all the models. The NRM appears to be consistent with the Scale Variation, while the NSRM has a positive uncertainty band like the DP model, but its width is much shorter.
\section{$t \bar t$ production}
\label{sec:top}
What has been said so far is valid for the cross section of Higgs production in gluon fusion at 13 TeV. We turn now our attention to another relevant process at the LHC: $t \bar t$ production in gluon fusion. There are some important differences between the $t \bar t$ and the Higgs case. First, the maximum available order for $t \bar t$ is NNLO, which means that we can at best extrapolate two residues. Then, the THU on the PDFs has a larger impact on $t \bar t$  than on the Higgs cross section~\cite{doweneed}. Finally, the soft approximation for $t \bar t$ is slightly worse than the Higgs case in predicting the first known orders~\cite{topn3loapprox}.

For what concerns our models, we can apply them to the $t \bar t$ hadronic cross section without significant modifications, apart for the ERM. In fact, the resummed cross section has the following form
\begin{equation}
\label{eq:finalratio}
\Sigma^{\rm res}(m^2,\xi,N)
=\hat\sigma^\text{LO}(m^2,\xi)
\sum_{\mathbf{I=1,8}}
\bar g_{\mathbf{I}}(\as)\exp\left[G_\mathbf{I}(N)\right]
+\mathcal{O}\left(\frac{1}{N}\right),
\end{equation}
where we can see the separation between the singlet and the octet components. The constants $\bar g_{\mathbf{I}}(\as)$ can be found in~\cite{topn3loapprox}. Accordingly, we need to modify Equation~\eqref{eq:expERM} by introducing two renormalon exponents:
\begin{equation}
E_{\mathbf 1}(\as) = \sum_{n=0}^\oo r_{{\mathbf 1},n}\,\, \as^{n+1}, \qquad E_{\mathbf 8}(\as) = \sum_{n=0}^\oo r_{{\mathbf 8},n}\,\, \as^{n+1}.
\end{equation}
Since we have the sum of two exponents, we extrapolate one residues from the first exponent and one from the second exponent. However, we don't have to worry about this in the ESRM since the model can be formulated with Eq.~\eqref{eq:ESRM}, where $\sigma^\text{soft}$ already includes the sum of the singlet and the octet contributions. Therefore, we are left with only one exponent for the renormalon part of the series.

\begin{figure}[p]
\centering
\includegraphics[width=0.8\textwidth]{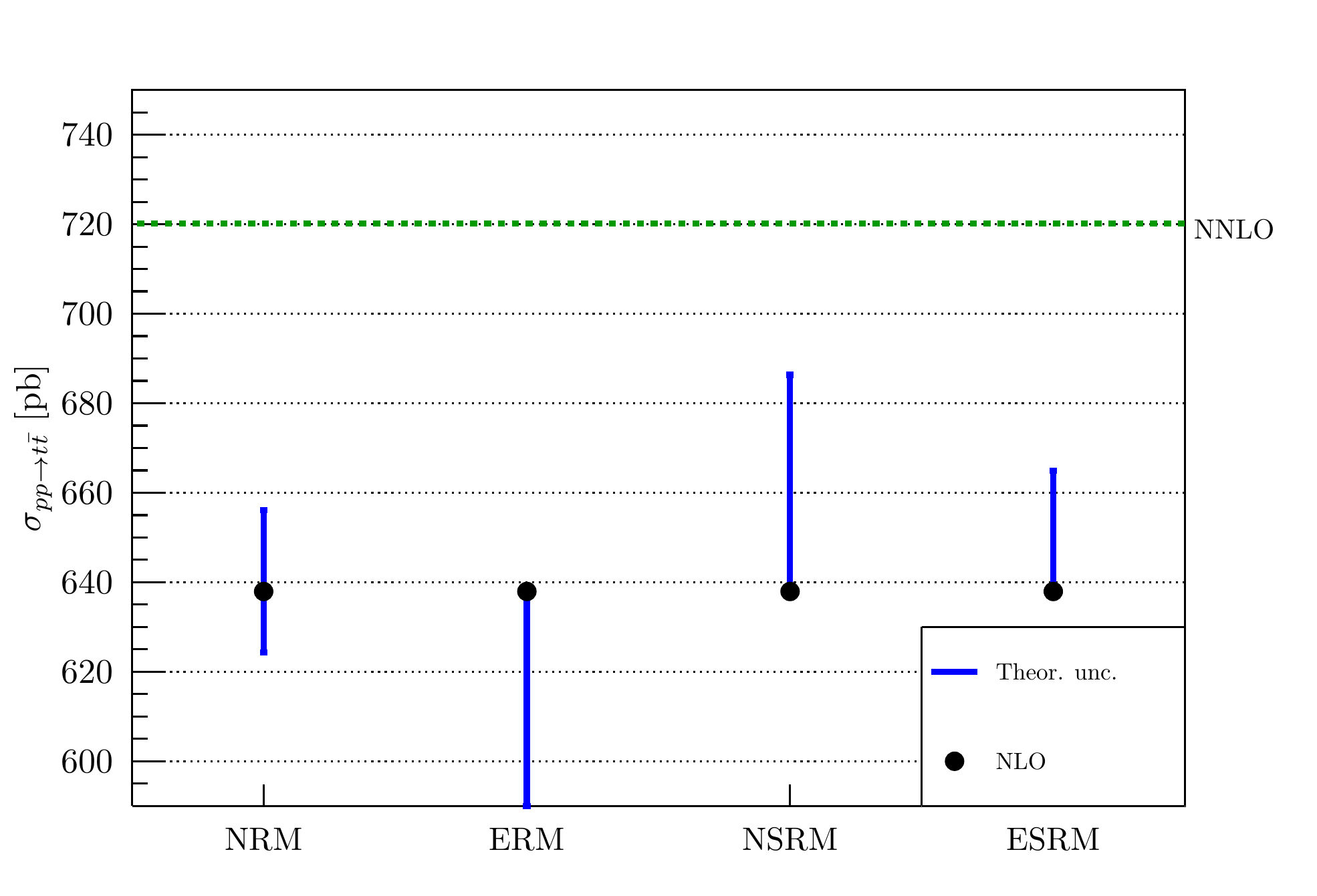}
\caption{Theoretical uncertainty bands on the NLO $t \bar t$ cross section (blue lines), computed pretending not to know the NNLO. The exact NNLO is displayed as a dashed green line. The fixed-order results and the resummed result are provided by \texttt{Top++}, the PDF set used is NNPDF3.0 with $\as = 0.118$.}
\label{fig:mynnlotop}
\end{figure}

We have tested the accuracy of our models at NLO, the penultimate available order. Results are shown in Fig.~\ref{fig:mynnlotop} (the equivalent of Fig.~\ref{fig:myn3lohiggs}). Once again, we decide to trust the Naive Models. In particular, the NSRM, our preferred choice for the Higgs cross section, is the most accurate in predicting the NNLO.

\begin{figure}[p]
\centering
\includegraphics[width=0.8\textwidth]{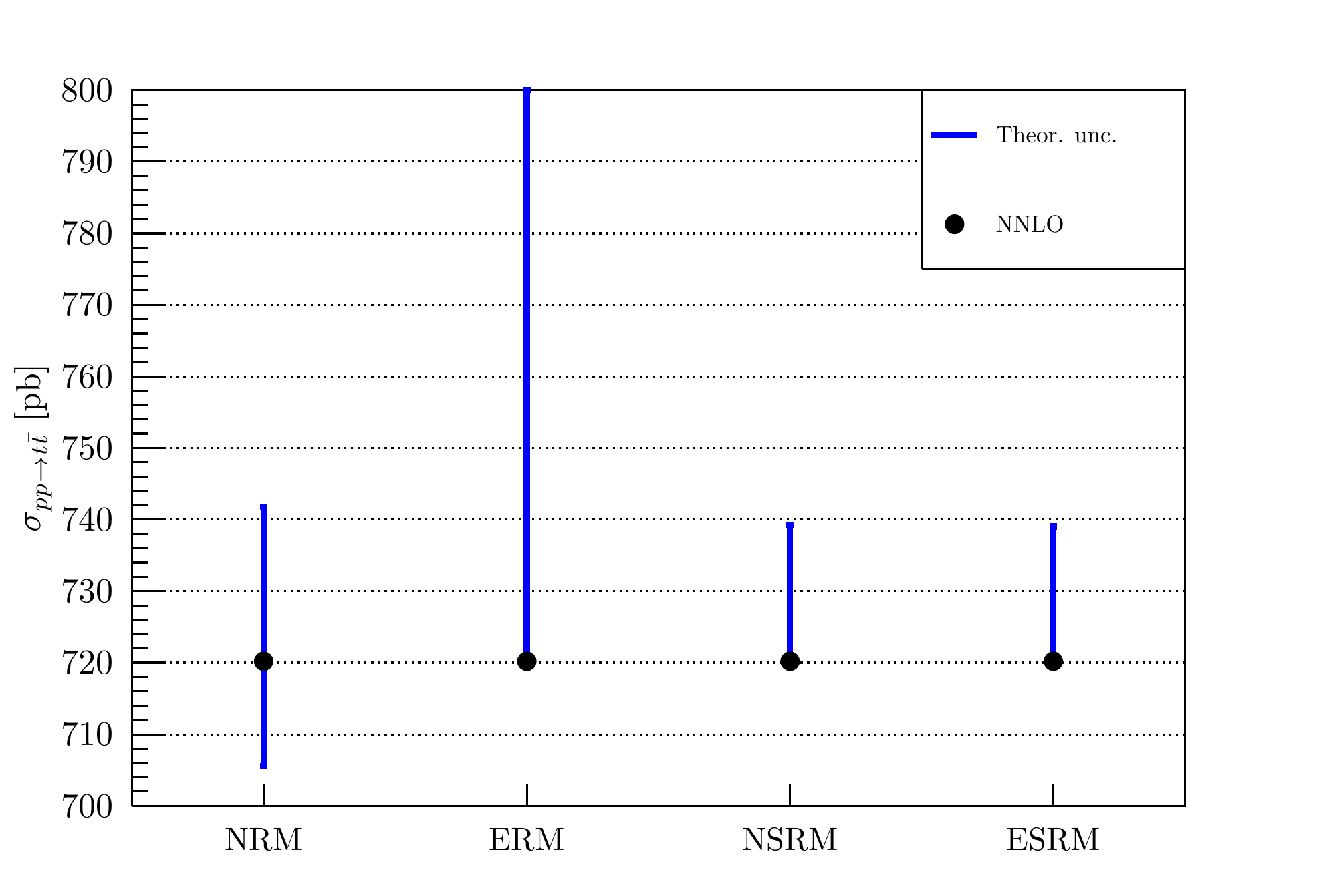}
\caption{Theoretical uncertainties on NNLO $t \bar t$ cross section (blue lines), predicted with each of our models. The NNLO is the last available known order for $t \bar t$ production. The settings are the same as the Figure above.}
\label{fig:mymodelcomp_top}
\end{figure}

Finally, Fig.~\ref{fig:mymodelcomp_top} shows the theoretical uncertainty bands on the NNLO. Comparing Fig.~\ref{fig:mymodelcomp_top} with Fig.~\ref{fig:mynnlotop}, we notice that once again the ERM model is unstable and changes drastically when a new order is included. The Naive Models have a behaviour which is very similar to the Higgs case. However, the NRM uncertainty band unexpectedly grows when a perturbative order is added. This does not necessarily mean that the NRM model does not converge (i.e. predicts higher and higher uncertainties). We recall that at NLO we only have one residue to determine, so we can expect an underestimation of the uncertainty since such little information is available.

We conclude that even for the $t \bar t$ case our preferred model is the most complete, accurate and convergent one: the Naive Soft Renormalon Model.

\phantomsection \chapter*{Conclusions} 
\thispagestyle{empty} 
\addcontentsline{toc}{chapter}{Conclusions}
In this thesis, we have studied the divergent behaviour of the perturbative series in QCD, taking as an example the Higgs cross section and the $t \bar t$ cross section. We have seen that there are three sources of divergence: the Landau pole, the renormalons and the instantons.  After introducing each of the sources, we have studied their phenomenological impact on the all-order cross section. 

In some theories, like the double-well potential in QM, the instanton divergence can completely cancel the divergence of the perturbative expansion. However, this is not the case of QCD because of the IR renormalons.
We found that the impact of the instanton divergence is totally negligible, as the partial sum $\sigma_n$ would start to increase significantly at $\bar n \sim 110$. 

The Landau pole divergence, related to the soft-gluon resummation,  occurs at high orders too ($\bar n \sim 33$). However, even if we don't have to worry about the divergence, we can construct models that combine soft contributions with renormalon contributions. We do so because we know that, accidentally, the expansion of the resummed cross section in the soft limit provides a good approximation of the exact known orders.

Finally, the dominant source of divergence in QCD is the renormalon divergence. Renormalons are poles in the Borel plane, which introduce an ambiguity in the definition of the asymptotic sum of the series. The divergence point for renormalons depends on the location of the leading pole, and in our case it turns out to be $\bar n \sim 13$.

After studying the divergent behaviour of the partial sum, we computed the asymptotic value of the sum of the series $\Sigma$ via the Borel integral method. We defined our estimate for the theoretical uncertainty as the difference between the last known order and $\Sigma$. However, the renormalon models allow a certain freedom in predicting $\Sigma$. In fact, the residues at the poles of the Borel transform cannot be computed theoretically and need to be extracted by comparison with the exact fixed-order results. The value of $\Sigma$ depends on what residues we decide to extract. The Borel transform has poles for each integer value of the real axis, but we can determine only 3 (Higgs) or 2 ($t \bar t$) residues. Therefore, combining the range of possible values for $\Sigma$ with their distance to the last known order, we can give an estimate to the theoretical uncertainty.

In particular, we have formulated 4 different models corresponding to 4 different ways of seeing the renormalon effect. First, we can simply assume that the perturbative expansion of the Higgs cross section presents the renormalon factorial divergence as it is (Naive Renormalon Model). Otherwise, we can assume that the renormalon corrections (basically identified with chains of bubble diagrams) are applied on the emitted soft gluons, and therefore the factorial behaviour is exponentiated (Exponent Renormalon Model). Finally, for both models we can consider the extension that takes into account the soft-gluon approximation. In other words, since we know that the soft-gluon approximation dominates at low orders, we ask that the renormalon divergence should cover the difference between the exact fixed-order series and its soft-gluon approximation. Those models are called Naive Soft Renormalon Model and Exponent Soft Renormalon Model.

We have compared the theoretical uncertainty predictions from our 4 models and we have found out that the uncertainty bands in the Exponent Models are bigger than in the Naive Models. Then we have investigated the accuracy of our models  by computing the uncertainty bands on the penultimate known order and checking if they predict the last known order correctly. In this investigation, we have noticed that the Exponent models are unstable for small variation of the initial parameters. So we have ruled them out and we have compared the NRM and the NSRM with the other already known models for theoretical uncertainties. Our models are compatible in magnitude with the conventional Scale Variation model. Actually, the NSRM, unlike the NRM and the Scale Variation, predicts a positive-only uncertainty, which one could expect since the perturbative series for the Higgs and the $t \bar t$ cross section appear to be monotonic.

Finally, we have compared theoretical uncertainties on the two different processes, finding similar results. Our preferred choice is the Naive Soft Renormalon Model, since 
\begin{itemize}
\item It is the most accurate in predicting the last known order
\item The uncertainty gets smaller when perturbative orders are added
\item It is stable for small variations of the initial conditions
\item It predicts a positive-only uncertainty, compatible with a monotonic partial sum.
\end{itemize}

In the NSRM, the Higgs and $t \bar t$ cross sections at the LHC at 13 TeV become
\begin{equation}
\sigma_{pp \to H} = 43.14^{+2.84\%}_{-0\%} \text{ pb},
\end{equation}
\begin{equation}
\sigma_{pp \to t \bar t} = 720.195^{+2.65\%}_{-0\%} \text{ pb}.
\end{equation}
\section*{Acknowledgments}
I would like to thank Marco Bonvini for the assistance in working with the Higgs codes, Claudio Muselli for providing the code for the $t \bar t$ soft-gluon approximation, Giuliano Giudici and Gherardo Vita for useful discussions on the subject of instantons and renormalons.

\end{document}